\documentclass[11pt]{article}
\pdfoutput=1
\usepackage{soul}
\usepackage{jheppub}
\usepackage{amsfonts}
\usepackage{amsmath}
\allowdisplaybreaks[4]         
\usepackage{amssymb}   
\usepackage{euscript}       
\usepackage{xcolor}          
\usepackage{tensor}     
\usepackage{caption}
\usepackage{subcaption}   
\usepackage{graphicx}
\usepackage[T1]{fontenc} % if needed
\usepackage{float}
\newcommand{\req}[1]{(\ref{#1})} %{Eq.\thinspace(\ref{#1})}  

\newcommand{\bea}{\begin{eqnarray}}
	\newcommand{\eea}{\end{eqnarray}}
\newcommand{\ba}{\begin{eqnarray}}
	\usepackage{braket}
	\newcommand{\ea}{\end{eqnarray}}
\newcommand{\nn}{\nonumber \\}

\newcommand{\beq}{\begin{equation}}
	\newcommand{\eeq}{\end{equation} }
\newcommand{\beqa}{\begin{eqnarray}}
	
	\newcommand{\eeqa}{\end{eqnarray}}
\newcommand{\beqar}{\begin{eqnarray*}}
	\newcommand{\eeqar}{\end{eqnarray*}}

\newcommand{\cT}{\mathcal{T}}
\def\th{\hbox{\char'336}}
\def\edth{\hbox{\char'360}}

\newcommand{\be}{\begin{equation}}
	\newcommand{\ee}{\end{equation}}
\newcommand{\diff}{\mathrm{d}}

\newcommand{\dv}{\\\notag &}
\newcommand{\dvv}{\right.\\\notag &\left.}

\newcommand{\dvvtag}{\right.\\ &\left.}

\renewcommand{\req}[1]{(\ref{#1})}

\newcommand{\R}{\mathcal{R}}

\newcommand{\D}{\mathcal{D}}
\newcommand{\E}{\mathcal{E}}

\newcommand{\cO}{\mathcal{O}}

\newcommand{\om}{\bar{\omega}}

\allowdisplaybreaks

\title{Teukolsky equation for near-extremal black holes beyond general relativity: near-horizon analysis}

\author[a]{Pablo A. Cano}
\author[b]{and Marina David}

\affiliation[a]{Departament de F\'isica Qu\`antica i Astrof\'isica, Institut de Ci\`encies del Cosmos Universitat de Barcelona, Mart\'i i Franqu\`es 1, E-08028 Barcelona, Spain}
\affiliation[b]{Instituut voor Theoretische Fysica, KU Leuven.
	Celestijnenlaan 200D, B-3001 Leuven, Belgium \vspace{0.1cm}}

\emailAdd{pablo.cano@icc.ub.edu}
\emailAdd{marina.david@kuleuven.be}
\date{\today}
\abstract{We study gravitational perturbations on the near-horizon region of extremal and near-extremal rotating black holes in a general higher-derivative extension of Einstein gravity. 
We find a decoupled modified Teukolsky equation that rules the gravitational perturbations and that separates into an angular and a radial equation.
 The angular equation leads to a deformation of the spin-weighted spheroidal harmonics, while the radial equation takes the same form as in Kerr except for a modification of the angular separation constants. 
We provide a detailed analysis of the corrections to these angular separation constants and find analytic results for axisymmetric modes as well as in the eikonal limit. 
As an application, we reproduce recent results that show that extremal Kerr black holes in higher-derivative gravity become singular under certain deformations and extend them by including parity-breaking corrections, which we show lead to the same effect. 
Finally, we obtain constraints on the form of the full modified Teukolsky radial equation by demanding that it has the right near-horizon limit. These results serve as an stepping stone towards the study of quasinormal modes of near-extremal black holes in higher-derivative extensions of GR.}

\begin{document} 
	\maketitle
	\flushbottom
	%%%%%%%%%%%%%%%%%%%%%%%%%%%%%%%%%%%%%
	%\setcounter{tocdepth}{2}
	%the line above sets the depth of the table of contents. {2} means it will display section and subsections only.
	%{\small
		%\setlength\parskip{-0.5mm} 
		%\tableofcontents
		%}

	\newpage
	\allowdisplaybreaks
	
	\section{Introduction}

One of the most relevant properties of a black hole is its quasinormal mode (QNM) spectrum \cite{Berti:2009kk,Konoplya:2011qq}, consisting of a set of modes with characteristic frequencies and damping times which determine the response of the black hole under perturbations. The quasinormal modes control the ringdown phase after a black hole binary merger, and are called to provide one of the cleanest tests of general relativity (GR) through gravitational-wave observations  \cite{Berti:2018vdi,Barack:2018yly}.  Furthermore, QNMs are key elements to assess the stability of a black hole solution.

 In recent years, there has been a growing effort in obtaining the quasinormal modes of black holes in a number of extensions of GR, \textit{e.g.} \cite{Cardoso:2009pk,Blazquez-Salcedo:2016enn,Cardoso:2018ptl,deRham:2020ejn,Cano:2020cao,Pierini:2021jxd,Wagle:2021tam,Srivastava:2021imr,Cano:2021myl,Pierini:2022eim,Silva:2024ffz}, motivated by the possibility of searching for beyond-GR physics with gravitational-wave observations \cite{Silva:2022srr,Franchini:2023eda,Maselli:2023khq}.  
However, the case of QNMs of rotating black holes --- the most relevant ones for observations --- has remained elusive.  In GR, the perturbations of Kerr black holes are described by the Teukolsky equation \cite{Teukolsky:1973ha}, a decoupled and separable equation stemming from the Petrov D character of the Kerr metric and the existence of a Killing tensor \cite{Walker:1970un}. The loss of these properties in extensions of GR prevents one from applying the Teukolsky equation and makes the study of perturbations a very challenging problem. Only very recently, with the development of generalized Teukolsky equations  \cite{Li:2022pcy,Hussain:2022ins,Cano:2023tmv,Wagle:2023fwl} that hold for arbitrary modifications of Einstein gravity, the computation of QNMs of modified Kerr black holes with substantial angular momentum has been possible \cite{Cano:2023jbk}.\footnote{See also \cite{Chung:2024ira,Chung:2024vaf} for a different approach based on spectral methods.}
In spite of this, the case of highly-spinning black holes --- close to extremality --- is still out of reach, since the methods developed so far rely on performing a series expansion in the angular momentum (of both the background metric and the modified Teukolsky equation), which breaks down at extremality. 

Near-extremal black holes, nevertheless,  are very interesting. Even in the case of Kerr black holes in GR, the computation of QNMs in the near-extremal regime is challenging and it reveals a rich and complex structure \cite{Yang:2012pj,Yang:2013uba}, with some modes becoming very long lived.  Thus,  corrections to GR, even if small, could render these black holes unstable. On a related front, it has recently been found in \cite{Horowitz:2023xyl} that static perturbations of extremal Kerr black holes in higher-derivative gravity can become singular at the horizon, provoking a divergence of tidal forces and a breakdown of the effective field theory (EFT) of gravity --- see \cite{Horowitz:2024dch} for the case of Kerr-Newman black holes. In addition, it was observed by \cite{Cano:2023jbk} that some of the higher-derivative corrections to the QNM frequencies grow very fast as the angular momentum of the black hole increases. This all points towards a possible enhancement of the effects of higher-derivative corrections near extremality,\footnote{Besides higher-derivative corrections, it has recently been shown that quantum effects become important for black holes very close to extremality \cite{Kapec:2023ruw,Rakic:2023vhv}. Which effect dominates first depends on the scale of the higher-derivative corrections and on the charge of the black hole \cite{Horowitz:2024dch}, although here we will only consider neutral black holes.} hence making these black holes perhaps the best candidates to test extensions of GR with gravitational waves.

In this paper, we take the first step towards the characterization of perturbations of near-extremal rotating black holes beyond GR. %In particular, we focus on the case of a general EFT extension of GR with higher-derivative corrections \cite{Endlich:2017tqa}, which is the only modification of Einstein's theory that preserves its degrees of freedom and its symmetries. 
We focus on the case of a general EFT extension of GR with higher-derivative corrections \cite{Endlich:2017tqa}.  Our interest on this theory lies on its generality and its physical motivation. Higher-derivative corrections are a general prediction of quantum gravity and are explicitly realized in string theory \cite{Gross:1986iv,Gross:1986mw,Bergshoeff:1989de}. The EFT approach allows us to capture any of these corrections in an agnostic way by introducing only a handful of free coupling constants. In fact, the EFT captures the most general modification of GR that preserves its massless degrees of freedom and symmetries (diffeomorphism and local Lorentz invariance). Thus, understanding the effects of these corrections on black hole physics is a relevant question.

The full metric describing near-extremal rotating black holes in these theories is still unknown, but here we make progress by focusing on the near-horizon region of these black holes --- which is itself a solution of the equations of motion --- and by studying the gravitational perturbations in this region. To this end we make use of the ``universal Teukolsky equations'' of \cite{Cano:2023tmv} applied to the near-horizon geometry of the higher-derivative-corrected black holes. We show that the resulting modified Teukolsky equation automatically separates into a radial and an angular equation. The angular equation is a deformation of the usual spin-weighted spheroidal harmonics equation that appears in the Kerr case, and it leads to a correction of the corresponding angular separation constants. 
These corrected angular separation constants enter into the radial equation, which we find to be otherwise unaffected by the higher-derivative corrections. This result allows us to obtain valuable information on the form of the full Teukolsky radial equation for near-extremal black holes in these theories and also permits us to reexamine the results of \cite{Horowitz:2023xyl} under the Teukolsky formalism.

A more detailed summary of our results and structure of the paper is as follows. 
In Section~\ref{sec:reviewofNHEK}, we review how to take the near-horizon limit of extremal Kerr black holes and examine the Teukolsky equation from the near-horizon perspective. We shift gears in Section~\ref{sec:NHEKwithHD}, by finding the corrected near-horizon solutions when higher derivative corrections are included. This sets us up to study the modified Teukolsky equation and its separation on the background of these solutions in Section~\ref{sec:perturbationtheory}. We then discuss in detail the corrections to the angular separation constants in Section~\ref{sec:correctionstodeltaB}. We prove that our results are robust as they are invariant under several types of gauge freedom and we provide numerical and analytical results for the corrected angular separation constants. We also study the eikonal limit of the angular separation constants, obtaining analytic formulas.
In Section~\ref{sec:divergence}, we reproduce the arguments of \cite{Horowitz:2023xyl} (on the divergence of tidal forces at the horizon) using our approach, and we extend them by including the effect of parity-breaking corrections. In Section~\ref{sec:connectionstofullTequation} we compare the modified Teukolsky equation obtained from the near-horizon region to the near-horizon limit of the full modified Teukolsky radial equation, which depends on an undetermined potential. This leads to a number of constraints that this potential must satisfy and therefore we obtain valuable information on the form of the full Teukolsky equation for near-extremal black holes. 
Finally, we give an overview of our results and discuss how they may be useful in a future computation of the QNMs of near-extremal black holes in Section~\ref{sec:conclusion}. 
More details on the Teukolsky equation and on the eikonal limit can be found in Appendices~\ref{app:Teukolsky} and~\ref{app:eikonal}, respectively.

	%%%%%%%%%%%%%%%%%%%%%%%%%%%%%%
	\section{Review: near-horizon limit of extremal Kerr} \label{sec:reviewofNHEK}
	%%%%%%%%%%%%%%%%%%%%%%%%%%%%%%
	%\subsection{Near-horizon limit of extremal Kerr}
	%%%%%%%%%%%%%%%%%%%%%%%%%%%%%%
	In this section, we review the near-horizon limit of the Kerr black hole solution in Einstein gravity. Along the way, we establish a few relevant concepts and we define the notation and conventions we use throughout the rest of the paper. 
	
	The Kerr metric is given by
	\begin{align}
		\begin{split}
		d \bar{s}^2= & -\left(1-\frac{2 M r}{\Sigma}\right) d t^2- \frac{4 M a r\left(1-x^2\right)}{\Sigma} d t d \phi +  \Sigma\left(\frac{d r^2}{\Delta}+\frac{d x^2}{1-x^2}\right) \\& + \left(r^2+a^2+\frac{2 M r a^2\left(1-x^2\right)}{\Sigma}\right)\left(1-x^2\right) d \phi^2,
		\end{split}
	\end{align}
	where
	\begin{align}\label{Delta}
		\Delta = r^2 - 2Mr + a^2, \qquad \Sigma = r^2 + a^2 x^2.
	\end{align}
	The coordinate $x=\cos\theta$ is to be taken from $[-1,1]$ and $\phi$ is $2\pi$-periodic. The metric depends on the mass $M$ and on the rotation parameter $a=J/M$, where $J$ is the angular momentum. For $a\le M$, there is an inner and an outer horizon, located at
	\begin{align}\label{rpmKerr}
		r_{\pm} = M \pm \sqrt{M^2 - a^2}\,.
	\end{align}
	Extremality is met when these two horizons merge into a degenerate horizon, which happens for $a=M$. At extremality, the horizon develops an infinite AdS$_{2}$ throat and as a result the near-horizon region itself becomes an exact solution of Einstein's equations. In order to zoom into the near-horizon metric, we introduce the coordinates
	\begin{align}\label{nhlimit}
		r =M + \epsilon \rho, \quad t= 2M^2 \frac{\tau}{\epsilon}, \quad \phi = \psi + \frac{M\tau}{\epsilon}\, ,
	\end{align}
	and take the limit $\epsilon\to 0$. This leads to the near-horizon extremal Kerr (NHEK) metric,
	
	\begin{align} \label{eq:NHExtremalMetric}
	 	ds^2 = M^2 \left\{ \left(1+x^2\right)\left(-\rho ^2  d\tau^2 + \frac{d\rho^2}{\rho ^2}\right) +\frac{\left(1+x^2\right)}{ \left(1-x^2\right)}dx^2 + 4\frac{ \left(1-x^2\right)}{(1+x^2)}(\rho  d\tau+d\psi)^2 \right\}\, .
	 \end{align}
	 The Newman-Penrose (NP) description of the Kerr metric is also necessary for the study of perturbations. To this end, we introduce a  frame consisting of four null vectors $l_{\mu}, n_{\mu}, m_{\mu}, \bar{m}_{\mu}$, with $l^{\mu}n_{\nu}=-m^{\mu}\bar{m}_{\mu}=-1$, such that 
	 %\begin{align}
	%	& l_{\mu} m^{\mu}=l_{\mu} \bar{m}^{\mu}=n_{\mu} m^{\mu}=n_{\mu} \bar{m}^{\mu}=0,
	%	\\ &
	%	l_{\mu} n^{\mu}=l^{\mu} n_{\mu}=-m_{\mu} \bar{m}^{\mu}=-m^{\mu} \bar{m}_{\mu}=1,
	%\end{align}
	 \begin{align*}
		\bar{g}_{\mu \nu}=-2l_{(\mu} n_{\nu)} +2 m_{(\mu} \bar{m}_{\nu)}\, .
	\end{align*}
	In the context of the Teukolsky equation, the appropriate choice of the NP frame is given by the Kinnersley tetrad \cite{Kinnersley:1969zza}, which expressed as 1-forms, reads
	\begin{align} \label{eq:tetraddown}
		\begin{aligned}
		l_{\mu}dx^{\mu} &= - dt + \frac{\Sigma }{\Delta }dr + a \left(1-x^2\right) d\phi\,,
		\\
		n_{\mu}dx^{\mu} &= \frac{1}{2}\left(-\frac{\Delta}{\Sigma}dt - dr + \frac{a \Delta  \left(1-x^2\right)}{\Sigma}d\phi\right)\,,
		\\
		m_{\mu} dx^{\mu} &= \frac{1}{\sqrt{2}\zeta^{*}}\left(-i a \sqrt{1-x^2} dt-\frac{\Sigma}{\sqrt{1-x^2}}dx+i \sqrt{1-x^2} \left(a^2+r^2\right)d\phi \right)\,,
		\\
		\bar m_{\mu} dx^{\mu} &= \frac{1}{\sqrt{2}\zeta^{}}\left(i a \sqrt{1-x^2} dt-\frac{\Sigma}{\sqrt{1-x^2}}dx-i \sqrt{1-x^2} \left(a^2+r^2\right)d\phi \right)\,,
		\end{aligned}
	\end{align}
	and where $\zeta = r - i a x$.  In the extremal near-horizon limit, this becomes
	\begin{equation}\label{limframe}
	\begin{aligned}
		l_{\mu}dx^{\mu} &= \frac{M^2 \left(x^2+1\right) \left(d\rho-\rho ^2 d\tau \right)}{\epsilon  \rho ^2}+O\left(1\right)\,,
		\\
		n_{\mu}dx^{\mu}&=- \frac{1}{2}\epsilon  \left(\rho ^2 d\tau+d\rho \right)+O\left(\epsilon ^2\right)\,,
		\\
		m_{\mu}dx^{\mu}&= -\frac{M \left(\left(x^2+1\right) dx+2 i \left(x^2-1\right) (\rho  d\tau+d\psi)\right)}{\sqrt{2}(1+ix)\sqrt{1- x^2}}+O\left(\epsilon\right)\,,
		\\
		\bar{m}_{\mu}dx^{\mu}&=
		-\frac{M \left(\left(x^2+1\right) dx-2 i \left(x^2-1\right) (\rho  d\tau+d\psi)\right)}{\sqrt{2}(1-ix)\sqrt{1-x^2}}+O\left(\epsilon\right)\, .
	\end{aligned}
	\end{equation}
	We note that $l_{\mu}\sim \epsilon^{-1}$ while $n_{\mu}\sim \epsilon$. However, one can achieve a finite limit by performing a Lorentz boost $(l_{\mu}, n_{\mu})\rightarrow (\epsilon l_{\mu}, \epsilon^{-1} n_{\mu})$. The resulting frame is relevant for the application of the Teukolsky formalism in the near-horizon metric.

A slightly different limit of the Kerr metric is the near-extremal near-horizon limit. The idea consists in taking the limit $a\rightarrow M$ at the same time as we zoom into the near-horizon region \cite{Bredberg:2009pv}. Specifically, we set
\begin{align}
		r =r_+ + \epsilon \rho, \quad t =2M^2 \frac{\tau}{\epsilon}, \quad \phi= \psi +  \frac{M\tau}{\epsilon}, \quad a=M - \frac{\rho_0^2}{8M}\epsilon^2\,,
\end{align}
and then we take the limit $\epsilon\to 0$. This yields the near-NHEK metric,
\begin{align}\label{nearnhek}
 	\begin{split}
	ds^2 &= M^2 \left\{ \left(1+x^2\right)\left(-\rho(\rho + \rho_0)  d\tau^2 + \frac{d\rho^2}{\rho(\rho + \rho_0)}\right) +\frac{\left(1+x^2\right)}{ \left(1-x^2\right)}dx^2 \right. \\ &  \left. \qquad\qquad\qquad \qquad\qquad\qquad \qquad\qquad\qquad+ 4\frac{ \left(1-x^2\right)}{(1+x^2)}((\rho+\tfrac{1}{2}\rho_0)  d\tau+d\psi)^2 \right\}\, .
	\end{split}
\end{align}
Observe that in this case the horizon at $\rho=0$ is not extremal. In fact, the parameter $\rho_0$ represents the Hawking temperature of the black hole as measured by a local observer, the precise relationship being
\begin{equation}\label{rhoT}
\rho_{0}=4\pi T_{\rm nh}\, ,
\end{equation}
where the temperature $T_{\rm nh}$ is defined as the inverse of the periodicity of the Euclidean time $\hat{\tau}=i\tau$. 
However, for an asymptotic observer the black hole still has a zero temperature, on account of the infinite redshift.  In fact 
\begin{equation}
T=\frac{\epsilon \rho_0}{8\pi M^2}\, ,
\end{equation}
which vanishes when $\epsilon\rightarrow 0$.

	 %%%%%%%%%%%%%%%%%%%%%%%%%%%%%%%
	 %%%%%%%%%%%%%%%%%%%%%%%%%%%%%%%
	 \subsection{The Teukolsky equation and its near-horizon extremal limit}\label{sec2:limitTeq}
	 %%%%%%%%%%%%%%%%%%%%%%%%%%%%%%%
	 %%%%%%%%%%%%%%%%%%%%%%%%%%%%%%%
The Teukolsky equations are a key tool in investigating gravitational perturbations of Kerr black holes. These involve the Teukolsky variables $\Psi_{0}$, $\Psi_{4}$, defined as the following components of the Weyl tensor,
 \begin{equation}
\Psi_0=C_{\alpha\beta\mu\nu}l^{\alpha}m^{\beta}l^{\mu}m^{\nu}\, ,\qquad \Psi_4=C_{\alpha\beta\mu\nu}n^{\alpha}\bar{m}^{\beta}n^{\mu}\bar{m}^{\nu}\, .
\end{equation}
Although these variables vanish on the Kerr background, their linear perturbations satisfy decoupled, second-order differential equations. Furthermore, these equations admit separable solutions of the form 
\begin{equation}\label{PsiSep}
\begin{aligned}
\Psi_0=&e^{-i\omega t+i m \phi}S_{+2}(x)R_{+2}(r)\, ,\\
\Psi_4=&e^{-i\omega t+i m \phi}\zeta^{-4}S_{-2}(x)R_{-2}(r)\, ,
\end{aligned}
\end{equation}
yielding the radial and angular Teukolsky equations,
	 \begin{align} \label{eq:Tequationgeneral}
	 	\Delta^{-s+1}\frac{d}{dr}\left[\Delta^{s+1}\frac{dR_s}{dr}\right] + V_s R_s &= 0\, ,\\\label{eq:Sequationgeneral}
		\frac{d}{dx}\left[(1-x^2)\frac{dS_{s}}{dx}\right]+\left[(a\omega)^2 x^2-2sa\omega x+ B_{lm}-\frac{(m+s x)^2}{1-x^2}\right]S_{s}&=0\, ,
	 \end{align}
	where the potential reads
	 \begin{align}
	 	\begin{split}\label{TPoten}
	 		V_s&=(a m)^2+\omega^2\left(a^2+r^2\right)^2-4 a m M r \omega+i s\left(2 a m(r-M)-2 M \omega\left(r^2-a^2\right)\right)\\&+\Delta\left(-a^2 \omega^2+s-B_{l m}+2 i r s \omega\right) \,.
	 	\end{split}
	 \end{align}
and where $s=\pm 2$. 
The  solutions of the angular equations are known as the spin-weighted spheroidal harmonics \cite{Berti:2005gp}, and denoted as $S^{lm}_{s}(x,a\omega)$, while $B_{lm}=B_{lm}(a\omega)$ are the angular separation constants.\footnote{Here we define $B_{lm}$ using the same conventions as \cite{Cano:2023tmv}, such that these constants are the same for $s=+2$ and $s=-2$.} The angular numbers $l$ and $m$ take the values $l=2,3,\ldots$, and $-l\le m\le l$. We also note that the variables $\Psi_0$ and $\Psi_4$ contain the same information \cite{Wald:1973wwa}, and thus the perturbations can be equivalently described by either the $s=+2$ or $s=-2$ equations. 

Let us then consider the near-horizon extremal limit of \eqref{eq:Tequationgeneral} and \eqref{eq:Sequationgeneral}. The key aspect to take into account is that the frequency $\omega$ must approach a particular value in order for the Teukolsky equation to have a well-defined limit. In order to see this, simply consider a perturbation in the NHEK metric that behaves as $\Psi\sim e^{-i\om \tau+i m\psi}$, where $\Psi$ denotes any perturbed variable such as $\Psi_0$ or $\Psi_4$.  Then, using \req{nhlimit}, this becomes
\begin{equation}
\Psi\sim e^{-i\om \tau+i m\psi}=e^{-i t\left(\frac{m}{2M}+\frac{\epsilon \om}{2M^2}\right)+i m\phi}\, .
\end{equation}
Therefore, the frequency of these near-horizon perturbations as measured in asymptotic time $t$ is
\begin{equation}\label{omegaas}
\omega=\frac{m}{2M}+\frac{\epsilon\, \om}{2M^2}\, . 
\end{equation}
As a consequence, in the limit $\epsilon \to 0$, all these near-horizon perturbations correspond to perturbations at the critical frequency for superradiance $\omega_{cr}=\Omega_{H}m$, where $\Omega_{H}=1/(2M)$ is the angular velocity of the horizon \cite{Bardeen:1999px}.

Thus, in order to derive the near-horizon limit of the Teukolsky equation, we plug \req{omegaas} and \req{nhlimit} into \req{eq:Tequationgeneral}. The result reads
\begin{equation}\label{nhTeuksec2}
\rho^{2-2s}\frac{d}{d\rho}\left[\rho^{2s+2}\frac{dR_{s}}{d\rho}\right]+\left[\om^2+2\rho\om(m- is)+\rho^2\left(s+\frac{7m^2}{4}-B_{lm}\right)\right]R_{s}=0\, .
\end{equation}
In the case of the angular equation, the only difference is that $a\omega$ is always fixed to $a\omega=m/2$.  
Finally, if one considers the near-horizon near-extremal limit instead, the result is
\begin{equation}
\begin{aligned}
\bar{\Delta}^{1-s}\frac{d}{d\rho}\left[\bar{\Delta}^{s+1}\frac{dR_{s}}{d\rho}\right]&+\left[\om^2+\frac{\rho_{0}^2m}{4}\left(m-2is\right)\right.\\
&\left.+(2\rho+\rho_0)(m- is)\om+\bar{\Delta}\left(s+\frac{7m^2}{4}-B_{lm}\right)\right]R_{s}=0\, ,
\end{aligned}
\end{equation}
where 
\begin{equation}
\bar{\Delta}=\rho(\rho+\rho_0)\, .
\end{equation} 
One can also study perturbations of the NHEK metric in global coordinates \cite{Dias:2009ex}, but we will not consider that case here since we are interested in the presence of a horizon. 

These results can also be obtained by applying the Teukolsky formalism on the NHEK or near-NHEK metrics directly, as we review in Section~\ref{sec:TeukNHEK}. The goal of the rest of the paper is to do the same analysis in the case of higher-derivative theories. Thus, we start by finding the corrections to the NHEK solution in the next section.

 %%%%%%%%%%%%%%%%%%%%%%%%%%%%%%%%%%%%%%%%%%%
 %%%%%%%%%%%%%%%%%%%%%%%%%%%%%%%%%%%%%%%%%%%
 \section{Near-horizon geometries in higher-derivative gravity} \label{sec:NHEKwithHD}
 %%%%%%%%%%%%%%%%%%%%%%%%%%%%%%%%%%%%%%%%%%%
 %%%%%%%%%%%%%%%%%%%%%%%%%%%%%%%%%%%%%%%%%%%
	  We are interested in studying higher-derivative corrections to Einstein gravity in the form of an EFT expansion,
	 \begin{align} \label{eq:actionwithcorrections}
	 S = \frac{1}{16\pi}\int d^4 x \Big[R + \ell^4 \mathcal{L}_{(6)} + \ell^6 \mathcal{L}_{(8)}+\ldots \Big]\, ,
	 \end{align}
	 where $\ell$ represents the length scale of new physics,  $\mathcal{L}_{(n)}$ represents the most general covariant Lagrangian with $2n$ derivatives of the metric, and we are setting $G=1$. Upon field redefinitions, there are no four-derivative corrections, there are only two independent corrections at cubic order \cite{Cano:2019ore}, 
	  \begin{align} \label{eq:L6}
	 	\mathcal{L}_{(6)}=\lambda_{\text{ev}} R_{\mu \nu}{ }^{\rho \sigma} R_{\rho \sigma}{ }^{\delta \gamma} R_{\delta \gamma}{ }^{\mu \nu}+\lambda_{\text{odd}} R_{\mu \nu}{ }^{\rho \sigma} R_{\rho \sigma}{ }^{\delta \gamma} \tilde{R}_{\delta \gamma}{ }^{\mu \nu}\, ,
	 \end{align}
	 and three independent invariants at quartic order \cite{Endlich:2017tqa},
	 \begin{align} \label{eq:L8}
	 	\mathcal{L}_{(8)}= \epsilon_1 \mathcal{C}^2+\epsilon_2\tilde{\mathcal{C}}^2+ \epsilon_3 \mathcal{C} \tilde{\mathcal{C}}\,,
	 \end{align}
	 where 
	  \begin{align}\label{eq:CCtilde}
	 	\mathcal{C}=R_{\mu \nu \rho \sigma} R^{\mu \nu \rho \sigma}, \quad \tilde{\mathcal{C}}=R_{\mu \nu \rho \sigma} \tilde{R}^{\mu \nu \rho \sigma}\,.
	 \end{align}
		 The corrections proportional to the dimensionless couplings $\lambda_{\text{ev}}, \epsilon_1$ and $\epsilon_2$ have even parity while $\lambda_{\text{odd}}$ and $\epsilon_3$ have odd parity as they involve a single dual Riemann tensor\footnote{Note that our definition of the dual Riemann tensor is different by a factor of $1/2$ from the one in \cite{Horowitz:2023xyl} and \cite{Reall:2019sah}.}
	 \begin{align}\label{dualRiem}
	 	\tilde{R}^{\mu \nu \rho \sigma}=\frac{1}{2} \epsilon^{\mu \nu \alpha \beta} R_{\alpha \beta}{ }^{\rho \sigma}\,.
	 \end{align}
The equations of motion are given by
\begin{align}\label{EFE}
\E_{\mu\nu}\equiv  G_{\mu\nu} -\ell^4 T^{(6)}_{\mu\nu} - \ell^{6}T^{(8)}_{\mu\nu}=0\,,
\end{align}
where the effect of the corrections is expressed in the form of effective stress-energy tensors $T_{\mu\nu}^{(n)}$ that read
\begin{align}
T_{\mu \nu}^{(n)}=-\tensor{P}{^{(n)}_{(\mu}^{\rho \sigma \gamma}} R_{\nu) \rho \sigma \gamma}+\frac{1}{2} g_{\mu \nu} \mathcal{L}_{(n)}-2 \nabla^\sigma \nabla^\rho P_{(\mu|\sigma| \nu) \rho}^{(n)}\,.
\end{align}
We define the tensor $P_{\mu\nu\alpha\beta}$ as the derivative of the full Lagrangian with respect to the Riemann tensor
\begin{equation}\label{defofP1}
	P_{\mu\nu\alpha\beta}=\frac{\partial \mathcal{L}}{\partial R^{\mu\nu\alpha\beta}}\,,
\end{equation}
and $P^{(n)}_{\mu\nu\alpha\beta}$ as the derivative of $\mathcal{L}_{(n)}$.
Applied explicitly for \eqref{eq:actionwithcorrections} we find\footnote{We note that, for Ricci-flat spacetimes, the three terms in the $\lambda_{\rm odd}$ term are equal. This is always the case when we are evaluating perturbative corrections to the Einstein's equations, since in that case one can evaluate the right-hand-side of \req{EFE} on a solution of Einstein gravity, \textit{i.e.}, a Ricci flat metric. }
\begin{align}
	& P_{\mu \nu \rho \sigma}^{(6)}=3 \lambda_{\mathrm{ev}} R_{\mu \nu}{}^{\alpha \beta} R_{\alpha \beta \rho \sigma}+\lambda_{\text {odd }}\left(R_{\mu \nu}{ }^{\alpha \beta} \tilde{R}_{\alpha \beta \rho \sigma}+R_{\mu \nu}{ }^{\alpha \beta} \tilde{R}_{\rho \sigma \alpha \beta}+R_{\rho\sigma}{ }^{\alpha \beta} \tilde{R}_{\mu\nu \alpha \beta}\right)\,, \\
	& P_{\mu \nu \rho \sigma}^{(8)}=4 \epsilon_1 \mathcal{C} R_{\mu \nu \rho \sigma}+2 \epsilon_2 \tilde{\mathcal{C}}\left(\tilde{R}_{\mu \nu \rho \sigma}+\tilde{R}_{\rho \sigma \mu \nu}\right)+\epsilon_3\left[2 \tilde{\mathcal{C}} R_{\mu \nu \rho \sigma}+\mathcal{C}\left(\tilde{R}_{\mu \nu \rho \sigma}+\tilde{R}_{\rho \sigma \mu \nu}\right)\right]\,.
\end{align}

Extremal rotating black hole solutions of the theory \req{eq:actionwithcorrections} are not known, as the only analytic solutions have been found using an expansion in the angular momentum \cite{Cardoso:2018ptl,Cano:2019ore} that breaks down at extremality. However, we do have access to the near-horizon geometries of these theories owing to their enhanced symmetry \cite{Chen:2018jed,Cano:2019ozf,Horowitz:2023xyl,Cano:2023dyg}.
	
In the effective theory  \req{eq:actionwithcorrections}, we can write an ansatz for the near-horizon geometry in the form
	  \begin{align}\label{NHEKmetricansatz}
	 	\begin{split}
	 		ds^2 = \mu^2 \Bigg\{ &(1+H_{1})\left(1+x^2\right)\left(-\rho^2 d\tau^2 + \frac{d\rho^2}{\rho^2}\right) +(1+H_{2})\frac{\left(1+x^2\right)}{ \left(1-x^2\right)}dx^2  \\ &  + 4(1+ H_{3})\frac{ \left(1-x^2\right)}{(1+x^2)}(d\psi +\Gamma \rho d\tau)^2 \Bigg\}\, ,
	 	\end{split}
	 \end{align}
where the functions $H_{i}(x)$, $i=1,2,3$, capture the deformation of the NHEK metric and $\Gamma$ is a constant introduced to ensure that $\psi$ is $2\pi$-periodic. The dimensionful parameter $\mu$ sets the size of the black hole, but in general it does not coincide with its mass $M$ anymore --- we find the relation below. It is useful to use the scale $\mu$ to define the dimensionless coupling constants
\begin{equation}\label{alphaq}
\alpha_{\rm ev}=\frac{\lambda_{\rm ev}\ell^4}{\mu^4}\, ,\quad  \alpha_{\rm odd}=\frac{\lambda_{\rm odd}\ell^4}{\mu^4}\, ,\quad \alpha_{1}=\frac{\epsilon_{1}\ell^6}{\mu^6}\, ,\quad \alpha_{2}=\frac{\epsilon_{2}\ell^6}{\mu^6}\, ,\quad \alpha_{3}=\frac{\epsilon_{3}\ell^6}{\mu^6}\, ,
\end{equation}	
which determine the size of the higher-derivative corrections. The EFT description is valid as long as $|\alpha_{\rm q}|\ll 1$.

The functions $H_{i}$ and the constant $\Gamma$ receive corrections at first order in the couplings, so we write
\begin{equation}
H_{i}=\sum_{\rm q}\alpha_{\rm q} H_{i,(\rm q)}\, ,\quad  \Gamma= 1+\sum_{\rm q}\alpha_{\rm q} \Gamma_{(\rm q)}\,.
\end{equation}	
We impose the functions $H_{i}$ to be regular (in particular, they should not diverge at $x=\pm 1$) while the value of $\Gamma$ is obtained by demanding the absence of conical defects. We find the following solution for the various $H_{i,q}$ functions, 
\begin{align}\notag
			H_{1,{\rm (ev)}} &= \frac{848 x^2}{7 \left(x^2+1\right)^2}\,,
			&H_{1,{\rm (odd)}}&= -\frac{96 x}{7 \left(x^2+1\right)^2}\,,
			&H_{1,{\rm (1)}} &= \frac{3 (416-105 \pi ) x^2}{\left(x^2+1\right)^2} \,,
			\\
			H_{1,{\rm (2)}}&= -\frac{9 (16+35 \pi ) x^2}{ \left(x^2+1\right)^2}\,
			&H_{1,{\rm (3)}} &= \frac{112 x}{5  \left(x^2+1\right)^2}\,,
			& &
\end{align}

\begin{align}\notag
			H_{2,{\rm (ev)}} &= -\frac{8 \left(25 x^{10}-10 x^8-510 x^6+2984 x^4-5339 x^2+482\right)}{7  \left(x^2+1\right)^6}\,,\\\notag
			H_{2,{\rm (odd)}}&= \frac{4 x \left(3 x^{10}+69 x^8+306 x^6+3230 x^4-10261 x^2+5309\right)}{7 \left(x^2+1\right)^6}\,,\\\notag
			H_{2,{\rm (1)}} &= -\frac{630 x \arctan x}{ \left(x^2-1\right)} + 
			\frac{1}{5  \left(x^2-1\right) \left(x^2+1\right)^9}\left(-942 x^{18}+2236 x^{16} \right. \\\notag & \left.+36924 x^{14} +114156 x^{12}+89984 x^{10} +1420500 x^8-3548988 x^6+2773860 x^4 \right. \\\notag & \left.-967890 x^2 +1575 \pi  (x^2+1)^7 (x^4+1)+80160\right) \,,\\\notag
			H_{2,{\rm (2)}}&=-\frac{630 x \arctan x}{ \left(x^2-1\right)} + \frac{1}{5  \left(x^2-1\right) \left(x^2+1\right)^9}\left(1575 \pi  \left(x^4+1\right) \left(x^2+1\right)^7 \right. \\\notag & \left. +2 \left(321 x^{18}+2182 x^{16}+6318 x^{14}+9702 x^{12}+87488 x^{10}-635670 x^8+1613394 x^6 \right. \right. \\\notag & \left. \left. -1572030 x^4+487935 x^2+360\right)\right)\,\\\notag
			H_{2,{\rm (3)}} &=- \frac{8 x}{5  \left(x^2+1\right)^9} \left(5 x^{16}+78 x^{14}+454 x^{12}+1462 x^{10}+1632 x^8+64394 x^6-239430 x^4 \right. \\& \left. +208498 x^2-37093\right)\,, 
\end{align}

\begin{align}\notag
			H_{3,{\rm (ev)}} &=\frac{2 \left(107 x^{12}+426 x^{10}+941 x^8+1628 x^6+1429 x^4+4858 x^2+83\right)}{7  \left(x^2+1\right)^6}\,,\\\notag
			H_{3,{\rm (odd)}}&=-\frac{4 x \left(3 x^{10}+21 x^8+114 x^6+350 x^4+1643 x^2-787\right)}{7 \left(x^2+1\right)^6}\,,\\\notag
			H_{3,{\rm (1)}} &= \frac{630 x \arctan x}{\left(x^2-1\right)}+\frac{1}{10 \left(x^2-1\right) \left(x^2+1\right)^9}\left(4 (192 x^{20}+295 x^{18} \right. \\\notag&  \left. -270 x^{16}+3810 x^{14}+18570 x^{12}+33632 x^{10}-99402 x^8+99102 x^6-48498 x^4\right. \\\notag & \left.-5767 x^2  -1664)-1575 \pi (x^2+1)^7 \left(x^6+x^4+3 x^2-1\right)\right) \,,\\\notag
			H_{3,{\rm (2)}}&=\frac{630 x \arctan x}{ \left(x^2-1\right)}-\frac{1}{10 \left(x^2-1\right) \left(x^2+1\right)^9}\left(1575 \pi  \left(x^6+x^4+3 x^2-1\right) \left(x^2+1\right)^7 \right. \\\notag & \left. +4 \left(672 x^{20}+4825 x^{18}+14430 x^{16}+21870 x^{14}+13350 x^{12}+6512 x^{10}-110502 x^8 \right. \right. \\\notag & \left. \left. +105042 x^6-47838 x^4-7177 x^2-1184\right)\right)\,\\\notag
			H_{3,{\rm (3)}} &= \frac{8 x}{5 \left(x^2+1\right)^9} \left(5 x^{16}+50 x^{14}+258 x^{12}+874 x^{10}+1772 x^8+14134 x^6\right.\\ & \left. -18258 x^4 +1326 x^2-161\right)\,, 
\end{align}
while for $\Gamma$ we find
	\begin{align}
		\Gamma_{({\rm ev})} = \frac{1}{7}, \quad \Gamma_{({\rm odd})}  = 0, \quad \Gamma_{({\rm 1})}  =  \frac{4864+1575 \pi }{20}, \quad \Gamma_{({\rm 2})} = \frac{4736+1575 \pi }{20}, \quad \Gamma_{({\rm 3})} = 0\, .
	\end{align}
We note that $\Gamma$ is only sensitive to the even-parity corrections. 

We observe that the ansatz \req{NHEKmetricansatz} has gauge freedom corresponding to an infinitesimal coordinate transformation $x\to x+h(x)$, which has the effect of shifting the $H_{i}$ functions according to 
\begin{align}\label{eq:gaugeh}
	H_{1} &\to H_{1} + \frac{2 x h(x)}{x^2+1}\,,
	\quad
	H_{2} \to H_{2} + 2 h'(x)-\frac{4 x h(x)}{x^4-1}\,,
	\quad
	H_{3} \to H_{3} + \frac{4 x h(x)}{x^4-1}\,. 
\end{align}
When computing physical quantities, it is important to check that they are invariant under this gauge transformation. We will use this in section~\ref{sec:generalpropertiesofthecoefficients} to test our results. 

Finally, the solution \req{NHEKmetricansatz} can straightforwardly be extended to a near-horizon near-extremal  metric by introducing the parameter $\rho_0$ as in \req{nearnhek}. Thus, we have 
\begin{align}\notag 
	ds^2 = \mu^2 &\left\{ (1+H_1)\left(1+x^2\right)\left(-\rho(\rho + \rho_0)  d\tau^2 + \frac{d\rho^2}{\rho(\rho + \rho_0)}\right) +(1+H_2)\frac{\left(1+x^2\right)}{ \left(1-x^2\right)}dx^2 \right. \\  & \left. + 4(1+H_3)\frac{ \left(1-x^2\right)}{(1+x^2)}\left(d\psi+ (\rho+\tfrac{1}{2}\rho_0) \Gamma  d\tau\right)^2 \right\}\, ,
	\label{nearnhek2}
\end{align}
where the $H_i$ functions and $\Gamma$ remain unchanged. The reason for this is that \req{nearnhek2} can be obtained from \req{NHEKmetricansatz} after a local coordinate transformation of the AdS$_{2}$ metric. However, the two spacetimes have different properties as that transformation is singular --- see \cite{Bredberg:2009pv}.  Let us note that the parameter $\rho_{0}$ has the same interpretation as in the near-NHEK metric \req{nearnhek} as it still is related to the near-horizon temperature according to \req{rhoT}.

%%%%%%%%%%%%%%%%%%%%%%%%%%%%%%%%
\subsection{Conserved charges}
%%%%%%%%%%%%%%%%%%%%%%%%%%%%%%%%
In order to fully characterize the black holes described by the near-horizon geometries \req{NHEKmetricansatz}, we must obtain their charges, namely, the angular momentum, mass and entropy.  We briefly review the necessary details for such a task.

For a given Killing vector $\xi^{\mu}$, there exists a conserved Noether current $\mathbf{J}_{\xi}$ given by\footnote{We define $\boldsymbol{\epsilon_{\mu_1\ldots \mu_{n}}}=\frac{1}{(4-n)!}\epsilon_{\mu_1\ldots\mu_{n}\nu_1\ldots\nu_{4-n}}dx^{\nu_{1}}\wedge\ldots \wedge dx^{\nu_{4-n}}$ and $\epsilon_{0123}=\sqrt{|g|}\epsilon^{0123}$.}  
\begin{equation} \label{eq: definition of Noether current}
	\boldsymbol{J}_{\xi}=\diff \mathbf{Q}_{\xi}+2\boldsymbol{\epsilon_{\mu}}\xi^{\nu}\tensor{\mathcal{E}}{^{\mu}_{\nu}}\,.
\end{equation}
The integration of the Noether charge 2-form $\mathbf{Q}_{\xi}$ at spatial infinity then yields the associated conserved charge. 
However, in the case of the angular momentum --- associated to the angular Killing vector $\partial_{\phi}$ --- one can show that the integration can be performed at any constant-time surface, including the horizon --- see \cite{Cano:2023dyg,Cassani:2023vsa} for recent examples and additional details.

Then the conserved charge associated to $\partial_{\phi}$ can be written as
\begin{equation} \label{eq: def of J}
	J=-2\int_{\Sigma}\left[\boldsymbol{\epsilon}_{\mu\nu}\left(\bar{P}^{\mu\nu\alpha\beta}\nabla_{\alpha}\xi_{\beta}+2\nabla_{\beta}\bar{P}^{\mu\nu\alpha\beta}\xi_{\alpha}\right)\right]\, ,\quad \xi=\partial_{\phi}\, ,
\end{equation}
where we have used $\bar P_{\mu\nu\alpha\beta}= P_{\mu\nu\alpha\beta}-P_{\mu[\nu\alpha\beta]}$ in order to ensure that $\bar P_{\mu\nu\alpha\beta}$ satisfies the Bianchi identity. We choose the surface $\Sigma$ to be the black hole horizon and once the dust settles, we find that the angular momenta receives corrections only through the even parity terms at first order in the couplings
\begin{align}
	J = \mu ^2+\frac{209 \lambda_{\rm{ev}}\ell^4}{7 \mu ^2}-\frac{(2368+2205 \pi ) \epsilon_1 \ell^6}{8 \mu ^4}-\frac{(5216+2205 \pi ) \epsilon_2 \ell^6}{8 \mu ^4}\,.
\end{align}
We are also interested in the Iyer-Wald entropy \cite{Wald:1993nt,Iyer:1994ys}
\begin{equation} \label{eq: IW entropy formula}
	S=-2\pi \int_{\mathcal{H}}d^{2}x\sqrt{h}P^{\mu\nu\alpha\beta}\epsilon_{\mu\nu}\epsilon_{\alpha\beta} ,
\end{equation}
where $h$ the determinant of the induced two-dimensional horizon metric and $\epsilon_{\mu\nu}$ is the binormal to the horizon, normalized via $\epsilon_{\mu\nu}\epsilon^{\mu\nu}=-2$. Likewise, the odd-parity corrections are zero at first order and we find
\begin{align}
	S &= 2 \pi  \mu ^2 +\frac{416 \pi  \lambda_{\rm{ev}}\ell^4}{7 \mu ^2} - \frac{18 \pi  (232+175 \pi ) \epsilon_1\ell^6}{5 \mu ^4}-\frac{18 \pi  (428+175 \pi ) \epsilon_2\ell^6}{5 \mu ^4} \,.
\end{align}
We can invert the relation and find the entropy in terms of the angular momentum $J$ to be
\begin{align}
	S&=
	2 \pi  J -\frac{2 \pi \lambda_{\rm{ev}}\ell^4}{7J} -\frac{\pi  (4864+1575 \pi ) \epsilon_1 \ell^6}{20 J^2}-\frac{\pi  (4736+1575 \pi ) \epsilon_2 \ell^6}{20J^2} \,,
\end{align}
which agrees with the results of \cite{Reall:2019sah}.

Finally, it is not possible to obtain the mass directly from the near-horizon geometry, but we can use the results of \cite{Reall:2019sah}, that provide the relation between mass and angular momentum of extremal black holes for the theories \req{eq:actionwithcorrections}. We then find the mass in terms of $\mu$ to be
\begin{align}
	M = \mu + 15 \frac{\lambda _{\rm{ev}} \ell^4}{\mu ^3} -\frac{\epsilon_1 \ell^6}{\mu^5}\frac{3 (4352+3675 \pi ) }{80}-\frac{\epsilon_2 \ell^6}{\mu^5}\frac{3 (9088+3675 \pi )}{80} \,.
\end{align} 
As expected, $\mu$ is equal to the ADM mass when the higher-derivative corrections are turned off.

%%%%%%%%%%%%%%%%%%%%%%%%%%%%%%%%%%%%%
\section{Perturbation theory} \label{sec:perturbationtheory}
%%%%%%%%%%%%%%%%%%%%%%%%%%%%%%%%%%%%%
Our goal now is to study the gravitational perturbations of the near-horizon metrics we have just obtained. We start by reviewing the case of perturbations of the NHEK metric \req{eq:NHExtremalMetric} in general relativity and move on afterwards to the case of higher-derivative gravity.

\subsection{Teukolsky equations for NHEK}\label{sec:TeukNHEK}
 In order to study perturbations in the Teukolsky formalism, we introduce the following Newman-Penrose tetrad $\tensor{\hat{e}}{^{a}_{\mu}}=\{\hat{l}_{\mu},\hat{n}_{\mu},\hat{m}_{\mu},\hat{\bar{m}}_{\mu}\}$ for the NHEK metric \req{eq:NHExtremalMetric}, 
\begin{equation}\label{frameNHEK}
\begin{aligned}
		\hat{l}_{\mu} dx^{\mu}&= \frac{M^2 \left(x^2+1\right) \left(d\rho-\rho ^2 d\tau \right)}{\rho ^2}\, .
		\\
	\hat{n}_{\mu} dx^{\mu}&= \frac{1}{2}\left(-\rho ^2 d\tau-d\rho \right)\, ,
		\\
		\hat{m}_{\mu} dx^{\mu}&= -\frac{M^2 \left(\left(x^2+1\right) dx+2 i \left(x^2-1\right) (d\psi+\rho  d\tau)\right)}{\sqrt{2}\zeta^{*}\sqrt{1- x^2}}\, ,
		\\
		\hat{\bar{m}}_{\mu} dx^{\mu}&=
		-\frac{M^2 \left(\left(x^2+1\right) dx-2 i \left(x^2-1\right) (d\psi+\rho  d\tau)\right)}{\sqrt{2}\zeta\sqrt{1-x^2}}\, ,
\end{aligned}
\end{equation}
which we obtained previously from the near-horizon limit of the Kinnersley tetrad \req{limframe}. Here 
\begin{equation}
\zeta=M(1-i x)\, ,
\end{equation}
and the hat in $\tensor{\hat{e}}{^{a}_{\mu}}$ denotes that these are background quantities. We then consider a perturbed tetrad
\begin{equation}
\tensor{e}{^{a}_{\mu}}=\tensor{\hat{e}}{^{a}_{\mu}}+\tensor{\delta e}{^{a}_{\mu}}\, ,
\end{equation}
which leads to a perturbed metric $\delta g_{\mu\nu}$, spin connection $\delta\gamma_{abc}$, and curvature tensor. We look at the Weyl invariants $\Psi_0$ and $\Psi_4$, defined as
\begin{equation}
\Psi_0=C_{\alpha\beta\mu\nu}l^{\alpha}m^{\beta}l^{\mu}m^{\nu}\, ,\qquad \Psi_4=C_{\alpha\beta\mu\nu}n^{\alpha}\bar{m}^{\beta}n^{\mu}\bar{m}^{\nu}\, .
\end{equation}
We note that the background value of these variables vanishes on the NHEK background, so to leading order in perturbation theory, $\Psi_0$ and $\Psi_4$ coincide with their linear perturbations, $\Psi_{0,4}=\delta\Psi_{0,4}$. In fact, the background geometry is of Petrov type D, like in the full Kerr solution, and in these circumstances one finds that the variables $\Psi_{0,4}$ satisfy the Teukolsky equations \cite{Teukolsky:1973ha}
\begin{equation}\label{eq:TeukO}
\mathcal{O}_{+2}^{(0)}(\Psi_0)=0\, ,\quad \mathcal{O}^{(0)}_{-2}(\Psi_4)=0\, ,
\end{equation}
where $\mathcal{O}_{s}^{(0)}$ represents the Teukolsky differential operator of spin weight $s$. For completeness, we provide the explicit form of these equations for a general type D background --- and their generalization in the case of higher-derivative gravity --- in the Appendix \ref{app:Teukolsky} [see \req{TeukOpDef}].

These equations are separable and thus we consider the following ansatz for the Teukolsky variables $\Psi_0$ and $\Psi_4$
\begin{equation}\label{PsiSep}
\begin{aligned}
\Psi_0=&e^{-i\om t+i m \psi}S_{+2}(x)R_{+2}(\rho)\, ,\\
\Psi_4=&e^{-i\om t+i m \psi}\zeta^{-4}S_{-2}(x)R_{-2}(\rho)\, .
\end{aligned}
\end{equation}
When evaluated on this ansatz, the Teukolsky equations then yield the following equations for $s=\pm 2$,
\begin{equation}\label{TeukSep}
\mathcal{O}_{s}^{(0)}(\Psi_{2-s})\propto R_{s} \D_{s,(0)}^2S_{s}+S_{s}\D_{\rho}^2R_{s}=0\, ,
\end{equation}
where $ \D_{s,(0)}^2$ and $\D_{\rho}^2$ are the operators

\begin{align}\label{Dsop}
 \D_{s,(0)}^2S_{s}&=\frac{d}{dx}\left[(1-x^2)\frac{dS_{s}}{dx}\right]+\left(\frac{m^2x^2}{4}-m s x-\frac{(m+s x)^2}{1-x^2}\right)S_{s}\, ,\\
\D_{\rho}^2R_{s}&=\rho^{-2s}\frac{d}{d\rho}\left[\rho^{2s+2}\frac{dR_{s}}{d\rho}\right]+\left(\frac{\om^2}{\rho^2}+\frac{2\om(m- is)}{\rho}+s+\frac{7m^2}{4}\right)R_{s}\, .
\label{Drop}
\end{align}
Thus, in order to solve \req{TeukSep} we look for eigenfunctions of the angular operator, yielding the equation
\begin{equation}\label{angulareq}
 \D_{s,(0)}^2S_{s}=-B_{lm} S_{s}\, .
\end{equation}
This is a particular case of the spin-weighted spheroidal harmonic equation \req{eq:Sequationgeneral} that we saw before.  In fact, it corresponds to setting $a\omega=m/2$ in \req{eq:Sequationgeneral}, consistent with the fact that all the perturbations in the near-horizon geometry correspond to a frequency $\omega=m/(2M)$. Therefore, the solutions of this equation are the  
spin-weighted spheroidal harmonics of the type $S_{s}^{lm}(x; m/2)$, but we will simply denote them by $S_{s}^{lm}(x)$, since these are the only spin-weighted spheroidal harmonics that we will be using from now on. The angular separation constants $B_{lm}$ are defined in a way such that they are the same for $s=+2$ and $s=-2$. For $m=0$ they read
\begin{equation}
B_{l0}=l(l+1)-4\, .
\end{equation}
For $m\neq 0$ there are no exact analytic results for $B_{lm}$ nor for $S_{s}^{lm}(x)$, but both of them can be obtained numerically, and here we make use of the Black Hole Perturbation Toolkit \cite{BHPToolkit} to this end. On the other hand, an important property of the spin-weighted spheroidal harmonics is their orthogonality, and we also normalize them so that they satisfy
\begin{equation}\label{ortho}
2\pi\int_{-1}^{1}dx S_{s}^{lm}(x)S_{s}^{l'm}(x)=\delta_{ll'}\, .
\end{equation}
Finally, when we use \req{angulareq} into \req{TeukSep}, we obtain the radial equation
\begin{equation}\label{Teuknhsec4}
\rho^{-2s}\frac{d}{d\rho}\left[\rho^{2s+2}\frac{dR_{s}}{d\rho}\right]+\left(\frac{\om^2}{\rho^2}+\frac{2\om(m-is)}{\rho}+s+\frac{7m^2}{4}-B_{lm}\right)R_{s}=0\, ,
\end{equation}
which precisely agrees with the limit of the full radial Teukolsky equation that we discussed in Section \ref{sec2:limitTeq}.

\subsection{Modified Teukolsky equations for higher-derivative gravity}\label{sec:modT}
Let us now study the perturbations of the near-horizon geometries \req{NHEKmetricansatz} in the case of higher-derivative gravity.  Our starting point is the ``universal Teukolsky equations'' of \cite{Cano:2023tmv}  --- see also \cite{Li:2022pcy,Hussain:2022ins} for other closely related approaches. We refer to \cite{Cano:2023tmv} for a detailed explanation of these equations and their applications to study black hole perturbations, and we also review their explicit form in  Appendix \ref{app:Teukolsky}. Here we explain the process of evaluating, decoupling and separating these equations.

Let $\alpha$ denote a parameter that characterizes the corrections to GR, and hence also controls the departure of the background metric from the NHEK metric --- for instance, $\alpha$ would be one of the couplings defined in \req{alphaq}.  Let us also consider a generic perturbation on top of the background geometries \req{NHEKmetricansatz}, which is described by a perturbed frame $\delta e^{a}$ and the corresponding perturbed spin connection $\delta \gamma_{abc}$ and perturbed Weyl tensor components $\delta \Psi_{n}$.
Then,  at first order in $\alpha$ and at linear order in perturbation theory, these universal Teukolsky equations  take the form
\begin{equation}
\begin{aligned}\label{eq:modTeukPsi}
\mathcal{O}^{(0)}_{+2}(\delta \Psi_{0})+\alpha  \mathcal{O}^{(1)}_{+2}(\delta \Psi_{n},\delta e^{a}, \delta \gamma_{abc})&=0\, ,\\
\mathcal{O}^{(0)}_{-2}(\delta \Psi_{4})+\alpha  \mathcal{O}^{(1)}_{-2}(\delta \Psi_{n},\delta e^{a}, \delta \gamma_{abc})&=0\, ,
\end{aligned}
\end{equation}
where $\mathcal{O}^{(0)}_{s}$ is the original Teukolsky operator as in \req{eq:TeukO}, and $\mathcal{O}^{(1)}_{s}$ is a differential operator that now acts on all the perturbed variables: $\delta \Psi_{n}$,  $\delta e^{a}$ and $\delta \gamma_{abc}$. The explicit form of $\mathcal{O}^{(1)}_{s}$ can be obtained by linearizing the full (non-linear) universal Teukolsky equation \req{eqn:universalteukolsky0} on a given background, yielding a very long and not illuminating expression. 
We note that the corrections to the Teukolsky equation arise due to two different contributions: dynamical corrections --- owed to extra terms in the (linearized) Einsteins's equations --- and background corrections --- owed to the modification of the near-horizon geometries.  Due to the latter, the corrections to the equation depend on the corrected background frame. In our case, in order to evaluate \req{eq:modTeukPsi} we use the natural generalization of the frame \req{frameNHEK} to the geometries \req{NHEKmetricansatz}, given by
\begin{equation}\label{frameNHEKcorrection}
\begin{aligned}
		\hat{l}_{\mu} dx^{\mu}&= \sqrt{1+H_{1}}\frac{\mu^2 \left(x^2+1\right) \left(d\rho-\rho ^2 d\tau \right)}{\rho ^2}\, .
		\\
		\hat{n}_{\mu} dx^{\mu}&=- \frac{1}{2} \sqrt{1+H_{1}}\left(\rho ^2 d\tau+d\rho \right)\, ,
		\\
		\hat{m}_{\mu} dx^{\mu}&= -\frac{\mu^2 \left(\sqrt{1+H_{2}}\left(x^2+1\right) dx+2 i \sqrt{1+H_{3}}\left(x^2-1\right) (d\psi+\Gamma\rho  d\tau)\right)}{\sqrt{2}\zeta^{*}\sqrt{1- x^2}}\, ,
		\\
		\hat{\bar{m}}_{\mu} dx^{\mu}&=
		-\frac{\mu^2 \left(\sqrt{1+H_{2}}\left(x^2+1\right) dx-2 i  \sqrt{1+H_{3}} \left(x^2-1\right) (d\psi+\Gamma \rho  d\tau)\right)}{\sqrt{2}\zeta\sqrt{1-x^2}}\, ,
\end{aligned}
\end{equation}
although we only need the frame to linear order in the correction functions $H_{i}$.

On the other hand, due to the presence of all the variables $\delta \Psi_{n},\delta e^{a}, \delta \gamma_{abc}$, these modified Teukolsky equations are not decoupled anymore.  The strategy to decouple them consists in expressing all these variables in terms of the Teukolsky variables alone, $\delta \Psi_{0,4}$. Furthermore, since the term $\mathcal{O}^{(1)}_{s}(\delta \Psi_{n},\delta e^{a}, \delta \gamma_{abc})$ already appears multiplied by $\alpha$ in \req{eq:modTeukPsi}, it suffices to find the relations between $\delta \Psi_{n},\delta e^{a}, \delta \gamma_{abc}$ and $\delta \Psi_{0,4}$ at zeroth-order in $\alpha$ --- that is, for Einstein gravity.
This is most easily done by reconstructing the metric perturbation on the NHEK geometry in terms of the Teukolsky variables, as we review next.

\subsubsection{Metric reconstruction}\label{sec:metricrec}
Let us then for the time being consider the case of perturbations of the NHEK metric in GR. 
It turns out that the metric perturbation can be written in terms of four Hertz potentials $\psi_{0,4}$, $\psi_{0,4}^*$ as \cite{Dolan:2021ijg}
\begin{equation}\label{hmunureconstructed}
 h_{\mu\nu}=-\frac{i}{3}\nabla_{\beta}\left[\zeta^4\nabla_{\alpha}\tensor{\mathcal{C}}{_{(\mu}^{\alpha}_{\nu)}^{\beta}}\right]-\frac{i}{3}\nabla_{\beta}\left[(\zeta^{*})^4\nabla_{\alpha}\tensor{\bar{\mathcal{C}}}{_{(\mu}^{\alpha}_{\nu)}^{\beta}}\right]\, ,
\end{equation}
where 
\begin{equation}
\begin{aligned}\label{Ctensor}
\mathcal{C}_{\mu\alpha\nu\beta}&=4\left( \psi_0 n_{[\mu}\bar{m}_{\alpha]}n_{[\nu}\bar{m}_{\beta]}+ \psi_4 l_{[\mu}m_{\alpha]}l_{[\nu}m_{\beta]}\right)\, ,\\
\bar{\mathcal{C}}_{\mu\alpha\nu\beta}&=4\left(\psi_0^{*} n_{[\mu}m_{\alpha]}n_{[\nu}m_{\beta]}+\psi_4^{*} l_{[\mu}\bar{m}_{\alpha]}l_{[\nu}\bar{m}_{\beta]}\right)\, .
\end{aligned}
\end{equation}
We remark that, since we allow for a complex metric perturbation, the potentials  $\psi_{0,4}^*$ are not really the complex conjugates of $\psi_{0,4}$ and they are regarded as independent variables. For the same reason, the conjugate Teukolsky variables $\delta\Psi_{0,4}^{*}$ are also independent from $\delta\Psi_{0,4}$ and we must consider their own equations. In the case of GR, the conjugate Teukolsky variables actually satisfy the same radial equations as $\delta\Psi_{0,4}$, but this is no longer the case for higher-derivative gravity, where one finds that the conjugate variables satisfy equations which are analogous, but different, to \req{eq:modTeukPsi}. Thus, in general it is necessary to include the four variables in the analysis of perturbations.

Coming back to Einstein gravity, we find that the Hertz potentials satisfy the Teukolsky equations, and therefore, we implement a separation of variables as in \req{PsiSep},
\begin{equation}\label{PsiSep2}
\begin{aligned}
\psi_0=&e^{-i\om t+i m \psi}S^{lm}_{+2}(x)\R_{+2}(r)\, ,\\
\psi_0^{*}=&e^{-i\om t+i m \psi}S^{lm}_{-2}(x)\R_{+2}^{*}(r)\, ,\\
\psi_4=&e^{-i\om t+i m \psi}\zeta^{-4}S^{lm}_{-2}(x)\R_{-2}(r)\, ,\\
\psi_4^{*}=&e^{-i\om t+i m \psi}(\zeta^{*})^{-4}S^{lm}_{+2}(x)\R_{-2}^{*}(r)\, ,
\end{aligned}
\end{equation}
where $S^{lm}_{s}(x)$ are the usual spin-weighted spheroidal harmonics that we discussed above. We observe that in the conjugate potentials the angular functions appear swapped, and we have $S^{lm}_{-2}(x)$ in $\psi_0^{*}$ and $S^{lm}_{+2}(x)$ in $\psi_4^{*}$. Also, we are using caligraphic $\R_{s}$ to distinguish these radial functions from those in Teukolsky variables, denoted $R_{s}$.  

The key result for the reconstruction of the metric perturbation is that the Teukolsky variables are simply proportional to the Hertz potentials,
\begin{equation}\label{metricweylrelation}
\delta\Psi_{2-s}=P_{s}\psi_{2-s}\, ,\quad \delta\Psi_{2-s}^{*}=P_{s}^{*}\psi_{2-s}^{*}\, .
\end{equation}
with fixed proportionality constants  $P_{s}$, $P_{s}^{*}$ that we show below. In order to obtain these relations, it is crucial to observe several relationships satisfied by the radial functions. First, we note that $\mathcal{R}_{s}$ and the conjugate radial variables $\mathcal{R}_{s}^{*}$ satisfy the same equations --- namely, \req{Teuknhsec4} --- and therefore these variables must be proportional,
\begin{equation}\label{conjugaterelation}
\begin{aligned}
\R_{+2}^{*}(r)&=q_{+2} \R_{+2}(r)\, ,\quad \R_{-2}^{*}(r)&=q_{-2} \R_{-2}(r)\, .
\end{aligned}
\end{equation}
The proportionality constants, $q_{s}$, which are a priori free parameters, turn out to determine the polarization of the perturbation.  For instance, one can see that the choices $q_{+2}=q_{-2}=\pm 1$ yield modes of definite parity under the antipodal symmetry $\phi\rightarrow\phi+\pi$, $x\rightarrow -x$  \cite{Cano:2023tmv}. Thus, for theories that preserve parity, the $q_{s}=+1$ and $q_{s}=-1$ modes naturally decouple. However, for theories that break parity, these modes are coupled, the polarization cannot be fixed a priori and the parameters $q_s$ must be determined by solving the equations for $\R_{s}$ and $\R_{s}^{*}$ simulteneously --- we come back to this later, in Section \ref{sec:generalpropertiesofthecoefficients}.

On the other hand, the radial functions with spin $s=+2$ and $s=-2$ are related by the Starobinsky-Teukolsky identities  \cite{teukolsky1972rotating,Starobinsky:1973aij,Chandrasekhar:1984siy,Fiziev:2009ud}
 \begin{equation}\label{STidentities}
\R_{-2}=C_{+2}\rho^4\left(\mathfrak{D}_{0}\right)^4\left(\rho^4 \R_{+2}\right)\, ,\quad
\R_{+2}=C_{-2}\left(\mathfrak{D}^{\dagger}_{0}\right)^{4}\R_{-2}\, ,
\end{equation}
where $\mathfrak{D}_{0}$ and $\mathfrak{D}^{\dagger}_{0}$ are the operators 

\begin{equation}
\mathfrak{D}_{0}=\partial_{\rho}+\frac{i\left(m \rho+\om\right)}{\rho^2}\, ,\quad \mathfrak{D}^{\dagger}_{0}=\partial_{\rho}-\frac{i\left(m \rho+\om\right)}{\rho^2}\, .
\end{equation}
The two proportionality constants $C_{\pm 2}$ satisfy a consistency condition obtained by applying the transformations twice,
\begin{equation}\label{Cproduct}
C_{+2}C_{-2}=\frac{1}{\mathcal{K}^2}\, ,
\end{equation}
where $\mathcal{K}$ is the radial Starobinsky-Teukolsky constant
\begin{equation}
\mathcal{K}^2=D_{2}^2+36 m^2\, ,
\end{equation}
and 
 \begin{equation}\label{eq:D2constant}
 \begin{aligned}
 D_{2}=\Bigg[&(B_{lm}+2)^2 (B_{lm}+4)^2-(B_{lm}+2) (B_{lm}+5) (3 B_{lm}-4) m^2\\
   &+\frac{3}{8} (B_{lm} (14+9 B_{lm})-50)
   m^4+\frac{9}{16} (1-3 B_{lm}) m^6+\frac{81 m^8}{256}\Bigg]^{1/2}\, ,
  \end{aligned}
 \end{equation}
is the angular Starobinsky-Teukolsky constant. 

By a direct evaluation of the Teukolsky variables $\delta\Psi_{0,4}$ and $\delta\Psi^{*}_{0,4}$ arising from the metric perturbation \req{hmunureconstructed}, and making use of the equations of motion of the Hertz potentials and the relationships \req{conjugaterelation} and \req{STidentities}, we obtain \req{metricweylrelation}, where the proportionality constants read
\begin{equation}\label{eq:Pconstants}
\begin{aligned}
P_{s}&=\frac{m}{4}+\frac{i s}{48 }\left(D_2 q_{s}- 2^s  q_{-s} C_{s}\mathcal{K}^2\right)\, ,\\
P_{s}^{*}&=\frac{m}{4}+\frac{i s}{48 q_{s}}\left(D_2-2^s C_{s} \mathcal{K}^2\right)\, ,
\end{aligned}
\end{equation}
with $s=\pm 2$.  This allows us to find the metric perturbation in terms of the Teukolsky variables by inverting \req{metricweylrelation}.

\subsubsection{Decoupled equations}
We now come back to the perturbations of near-horizon geometries in higher-derivative gravity.
With the results from the last subsection, we are now ready to rewrite the equations \req{eq:modTeukPsi} --- and their NP conjugates --- using only the perturbed Teukolsky variables.

We start by considering a general separation of variables for all the Teukolsky variables
\begin{equation}
\begin{aligned}
\delta\Psi_0=&e^{-i\om t+i m \psi}S_{+2}(x)R_{+2}(r)\, ,\\
\delta\Psi_0^{*}=&e^{-i\om t+i m \psi}S^{*}_{-2}(x)R^{*}_{+2}(r)\, ,\\
\delta\Psi_4=&e^{-i\om t+i m \psi}\zeta^{-4}S_{-2}(x)R_{-2}(r)\, ,\\
\delta\Psi_4^{*}=&e^{-i\om t+i m \psi}\left(\zeta^{*}\right)^{-4}S^{*}_{+2}(x)R^{*}_{-2}(r)\, .
\end{aligned}
\end{equation}
Here $S_{s}$ and $S_{s}^{*}$ are not assumed to be the spin-weighted spheroidal harmonics; they are arbitrary functions. However, they will take form of the spin-weighted spheroidal harmonics plus an order $\alpha$ correction, $S_{s}=S_{s}^{lm}+\mathcal{O}(\alpha)$,  $S_{s}^{*}=S_{s}^{lm}+\mathcal{O}(\alpha)$. Then, in order to evaluate \req{eq:modTeukPsi}, we need to determine the metric perturbation that gave rise to these perturbed Teukolsky variables. In fact, it suffices to find the metric perturbation in Einstein gravity, since $\mathcal{O}(\alpha)$ terms in the process of metric reconstruction will actually result in $\mathcal{O}(\alpha^2)$ terms in the equations  \req{eq:modTeukPsi}. Thus, we use \req{hmunureconstructed} with \req{metricweylrelation} and \req{eq:Pconstants}. Then, from the metric perturbation we obtain the perturbed frame $\delta e^{a}$, spin connection $\delta \gamma_{abc}$ and the rest of the Weyl variables $\delta\Psi_{n}$, and we plug them in \req{eq:modTeukPsi}. The resulting equations read
\begin{equation}
\begin{aligned}\label{TeukSep2}
R_{s}\left( \D_{s,(0)}^2S_{s}+\alpha \D_{s,(1)}S_{s}\right)+S_{s}\D_{\rho}^2R_{s}&=0\, ,\\
R_{s}^{*}\left( \D_{s,(0)}^2S_{s}^{*}+\alpha \D_{s,(1)}^{*}S_{s}^{*}\right)+S_{s}^{*}\D_{\rho}^2R_{s}^{*}&=0\, ,
\end{aligned}
\end{equation}
 where $\D_{s,(0)}^2$ and $\D_{\rho}^2$ are the operators \req{Dsop} and \req{Drop}, respectively. The term $\alpha \D_{s,(1)}$ represents a correction to the operator $\D_{s,(0)}^2$, thus giving rise to a modification of the angular equations, 
 \begin{equation}\label{Scorrected}
 \begin{aligned}
 \D_{s,(0)}^2S_{s}+\alpha \D_{s,(1)}S_{s}=-\left(B_{lm}+\delta B_{s,lm}\right)S_{s}\, ,\\
  \D_{s,(0)}^2S_{s}^{*}+\alpha \D_{s,(1)}^{*}S_{s}^{*}=-\left(B_{lm}+ \delta B_{s,lm}^{*}\right)S_{s}^{*}\, ,
 \end{aligned}
 \end{equation}
where the total angular separation constant receives an order-$\alpha$ correction, $\delta B_{s,lm}$ or $\delta B^{*}_{s,lm}$, which has to be determined as a part of the eigenvalue problem.\footnote{For convenience, we are absorbing the parameter $\alpha$ in the definition of $\delta B_{s,lm}$ and $\delta B^{*}_{s,lm}$.} Since at zeroth order in $\alpha$ the angular functions are nothing but the spin-weighted spheroidal harmonics, we can always assume the spheroidal equations in order simplify the term $\alpha \D_{s,(1)}S_{s}$. In fact, this allows us to always reduce it to a first-order operator, 

\begin{equation}\label{DiffsDef}
\D_{s,(1)}S_{s}=U_{s} S_s+V_{s} \frac{dS_{s}}{dx}\, ,\quad \D_{s,(1)}^{*}S^{*}_{s}=U^{*}_{s}S^{*}_s+V^{*}_{s}\frac{dS^{*}_{s}}{dx}\, ,
\end{equation}
where $U_{s}$, $V_{s}$ are functions of $x$ that depend on the theory and on the parameters $q_{s}$ and $C_{s}$. We have obtained explicitly these functions, which take very lengthy expressions. 

Interestingly, the radial equations --- which are obtained by plugging \req{Scorrected} into \req{TeukSep2} --- do not receive any explicit corrections, and the only modifications appear through the correction of the angular separation constants,
\begin{equation}\label{eq:RadialCorrected}
\begin{aligned}
\rho^{-2s}\frac{d}{d\rho}\left[\rho^{2s+2}\frac{dR_{s}}{d\rho}\right]+\left(\frac{\om^2}{\rho^2}+\frac{2\om(m-is)}{\rho}+s+\frac{7m^2}{4}-B_{lm}- \delta B_{s,lm}\right)R_{s}&=0\, ,\\
\rho^{-2s}\frac{d}{d\rho}\left[\rho^{2s+2}\frac{dR_{s}^{*}}{d\rho}\right]+\left(\frac{\om^2}{\rho^2}+\frac{2\om(m-is)}{\rho}+s+\frac{7m^2}{4}-B_{lm}- \delta B_{s,lm}^{*}\right)R_{s}^{*}&=0\, .
\end{aligned}
\end{equation}
However, these have some interesting consequences that we explore in section \ref{sec:divergence}. 

Importantly, the consistency of these radial equations implies that all the angular separation constants must, in fact, be equal, 
 \begin{equation}\label{equalB}
 \delta B_{+2,lm}=\delta B_{+2,lm}^{*} =\delta B_{-2,lm}=\delta B_{-2,lm}^{*}\equiv \delta B_{lm}\, .
 \end{equation}
To understand the origin of this condition, we must bear in mind that we are implicitly looking for quasinormal modes. These satisfy fixed boundary conditions at the horizon and at infinity, and they only occur for specific values of the frequency. Since a single perturbation is described by four radial equations, the four of them must give rise to the same spectrum of QNM frequencies in order for the full solution to be consistent. This means that the four equations must be identical, modulo changes of variable. We see that the equations for $R_{s}$ and $R_{s}^{*}$ are in fact identical if and only if $\delta B_{s,lm}=\delta B_{s,lm}^{*}$. On the other hand, one can apply the Teukolsky-Starbinsky identities \req{STidentities} to relate the $s=+2$ and $s=-2$ equations, and in this case $\delta B_{s,lm}=\delta B_{-s,lm}$ appears as the condition for one equation to exactly map into the other (hence guaranteeing that both share the same spectrum).  The consistency conditions \req{equalB} are crucial in order to completely determine the gravitational perturbation, as they fix the polarization (controlled by the parameters $q_{s}$), as we explain in the next section. 

Finally, we have repeated all the computations for the case of perturbations of near-extremal near-horizon metrics \req{nearnhek2}, and we have found that the angular equation remains unchanged, while the radial equation is again the same as in Kerr, except for the correction to $B_{lm}$, namely
\begin{equation}\label{Tnext}
\begin{aligned}
\bar{\Delta}^{1-s}\frac{d}{d\rho}\left[\bar{\Delta}^{s+1}\frac{dR_{s}}{d\rho}\right]&+\left[\om^2+\frac{\rho_{0}^2m}{4}\left(m-2is\right)\right.\\
&\left.+(2\rho+\rho_0)(m- is)\om+\bar{\Delta}\left(s+\frac{7m^2}{4}-B_{lm}-\delta B_{lm}\right)\right]R_{s}=0\, .
\end{aligned}
\end{equation}
 
 %%%%%%%%%%%%%%%%%%%%%%%%%%%%%%%%%%%%%%%%%%%%%%%%%%%%%%%%%%%%%%%%%%%%%%%%
 %%%%%%%%%%%%%%%%%%%%%%%%%%%%%%%%%%%%%%%%%%%%%%%%%%%%%%%%%%%%%%%%%%%%%%%%
 \section{Corrections to the angular separation constants} \label{sec:correctionstodeltaB}
 %%%%%%%%%%%%%%%%%%%%%%%%%%%%%%%%%%%%%%%%%%%%%%%%%%%%%%%%%%%%%%%%%%%%%%%%
 %%%%%%%%%%%%%%%%%%%%%%%%%%%%%%%%%%%%%%%%%%%%%%%%%%%%%%%%%%%%%%%%%%%%%%%%
In this section we analyze in detail the corrections to the angular separation constants, $\delta B_{lm}$, which play a relevant role in the radial equation \req{Tnext}. 
In order to find $\delta B_{lm}$, we need to solve the eigenvalue equations \req{Scorrected}. Since the undeformed operator $ \D_{s,(0)}^2$ is Hermitian and the spin-weighted spheroidal harmonics form an orthonormal basis of functions, we can use standard arguments of eigenvalue perturbation theory in order to find $\delta B_{lm}$. In fact, we can expand the correction to the angular functions in terms of the spheroidal harmonics,
\begin{equation}
S_{s}(x)=S_{s}^{lm}(x)+\alpha\sum_{l'} c_{l'}S_{s}^{l'm}(x)+\mathcal{O}(\alpha^2)\, ,
\end{equation}
and then, projecting \req{Scorrected} on $S_{s}^{lm}$ and taking into account the orthogonality relations \req{ortho}, yields 
 \begin{equation}
  \begin{aligned}
 \delta B_{s,lm}&=-2\pi\alpha \int_{-1}^{1}dx  \, S_{s}^{lm}(x) \D_{s,(1)}S_{s}^{lm}(x)\,,
 \end{aligned}
 \end{equation}
 and a similar expression for $\delta B_{s,lm}^{*}$
 Using \req{DiffsDef} and integrating by parts,\footnote{In doing this we observe that $V_s(x)=0$ at $x=\pm 1$ so no boundary terms are generated during the integration by parts.} we can always reduce these expressions to integrals of the form 
  \begin{equation}\label{deltaBf}
 \delta B_{s,lm}=2\pi \int_{-1}^{1}dx  \, \left(S_{s}^{lm}(x)\right)^2 f_{s}(x)\, ,
 \end{equation}
 where 
 \begin{equation}
 f_{s}=\alpha\left(-U_{s}+\frac{1}{2}V_{s}'\right).
 \end{equation}
 A similar formula holds for $ \delta B_{s,lm}^{*}$. The explicit expressions for $f_{s}(r)$ are too long to be displayed in the text, but we provide them in an ancillary Mathematica notebook.

 \subsection{General properties of the coefficients} \label{sec:generalpropertiesofthecoefficients}
 Before computing explicitly the value of the coefficients $\delta B_{lm}$, we make sure that their definition is robust as they satisfy several consistency checks. 
 
First of all,  we recall that there is gauge freedom in the metric functions $H_{i}(x)$ of the background near-horizon metric, given by the choice of a free function $h(x)$ --- see \req{eq:gaugeh}. The integrand  of $\req{deltaBf}$ depends explicitly on this gauge function, but the coefficient  $\delta B_{lm}$ should be independent of it. Here we show that this is indeed the case.  
The gauge function $h(x)$ has the following effect on the integral \req{deltaBf}
\begin{equation}\label{eq:integralgaugeterm}
	\delta B_{s,lm}=  2\pi \int_{-1}^{1}dx  \, \left(S_{s}^{lm}(x)\right)^2 \left[f_{s}(x)+\Delta f_{s}(h(x))\right]\, ,
\end{equation}
where $\Delta f_s(h(x))$ is the gauge term, which reads
\begin{equation}\label{eq:gaugeterm}
\begin{aligned}
\Delta f_s(h(x))&=-\frac{1}{2} \left(1-x^2\right) h^{(3)}(x)+\frac{1}{2(1-x^2)}\left(m^2 \left(x^4-x^2+4\right)
	\right. \\& \left.
	- 4 s m x \left(x^2-3\right) + 16 x^2- 4 - 4 B_{lm} (1-x^2) \right) h'(x) - \frac{1}{2(1-x^2)^2}\left(4B_{lm} x \left(1-x^2\right)
	 \right. \\& \left. 
	- m^2 x \left(x^2+7\right) + 2 s m \left(x^4-6 x^2-3\right) - 8 \left(2 x^3+x\right)\right)h(x)\, .
\end{aligned}
\end{equation}
Note that this term is of third order in derivatives of $h(x)$. We now show that in fact the integral of $\Delta f_{s}(h(x))$ in \eqref{eq:integralgaugeterm} vanishes and does not affect the shift in the angular separation constants. We first consider integrating by parts so that the integrand is only dependent on $h(x)$ and $h'(x)$ and no other derivatives on $h(x)$. In this way, we swap the derivatives of $h(x)$ to $S^{lm}_{s}(x)$. The result is
\begin{align}
	 &2\pi \int_{-1}^{1}dx  \, \left(S_{s}^{lm}(x)\right)^2 \Delta f_{s}(h(x))=g_s(h(x)) \Big\rvert^{1}_{-1}\nonumber  \\&   -2\pi \int_{-1}^{1}dx  \, \left(h'(x)S^{lm}_{s}(x)+\frac{2x }{1-x^2}h(x)S^{lm}_{s}(x)  -2h(x)\frac{dS^{lm}_{s}(x)}{dx}   \right)\left(\mathcal{D}^{2}_{s,(0)}+B_{lm}\right)S^{lm}_{s}(x)\, .
\end{align}
Now the integral in the second line is proportional to the differential equation of $S^{lm}_{s} (x)$, as defined in \req{angulareq}, and therefore it is identically zero. Let us then take a look at the boundary term $g_{s}(h(x))$ arising from the integration by parts, which takes the form  
\begin{align} \label{eq:boundaryterm}
	g_s(h(x)) &=- h(x)S_{s}^{lm}(x)\left(\mathcal{D}^{2}_{s,(0)}+B_{lm}\right)S^{lm}_{s}(x)
	\nonumber \\& -
	\left(1-x^2\right)\left[-S_{s}^{lm}(x)\frac{d}{dx}\left(h(x)\frac{dS_{s}^{lm}(x)}{dx}\right)
	+ h(x) \left(\frac{dS_{s}^{lm}(x)}{dx}\right)^2 + \frac{1}{2}\left(S_{s}^{lm}(x)\right)^2 h''(x)\right]\nonumber\\
	& - \left(S_{s}^{lm}(x)\right)^2 \left( x h'(x)+\frac{\left(1+x^2\right) }{1-x^2}h(x)\right)\, .
\end{align}
The first term automatically yields zero since it is again proportional to the equation of  $S^{lm}_{s} (x)$ \req{angulareq}. As long as the functions $h(x)$ and $S_{s}^{lm}(x)$ are regular at $x=\pm 1$, the term in the second line also vanishes as it is proportional to $(1-x^2)$. Finally, we take a closer look at the term in the third line. For general values of $l$ and $m$, the function $S_{s}^{lm}(x)$ does not necessarily vanish at $x=\pm 1$. However, one can see that, as long as $h(x)$ is a differentiable function such that 
\begin{align} \label{eq:hcondition}
	h(1) = h(-1)=0\, ,
\end{align}
then the third line in \req{eq:boundaryterm} vanishes at $x=\pm 1$.  Note that the function $h(x)$ corresponds to an infinitesimal change of coordinates $x'=x+h(x)$, and thus the condition \eqref{eq:hcondition} has a very intuitive reason to be: it guarantees that the new coordinate $x'$ also takes values in the interval $[-1,1]$. 

Therefore, we have shown that $\delta B_{s,lm}$ is independent of the gauge choice of the background metric. 
An interesting application of this result is that we can try to choose $h(x)$ to get a simpler expression for the integrand of $\delta B_{s,lm}$. We show this explicitly in Section~\ref{sec:eikonal}. 

Now, our construction contains another type of gauge freedom, in this case concerning the form of the metric perturbations. We recall that the metric perturbation depends on the choice of one of the Starobinsky-Teukolsky constants $C_{s}$, which is essentially a gauge parameter. For instance, $C_{-2}=0$ would correspond to the so-called outgoing radiation gauge, while $C_{+2}=0$ would be the ingoing radiation gauge, and in general we can have a combination of both. However, the final result should be independent of the choice of  the constants $C_{s}$. Alongside these, the expression of $f_{s}$ also depends on the polarization parameters $q_{s}$ that we introduced in Section~\ref{sec:metricrec}.
When one performs the integrals \req{deltaBf}, one in general finds that the result depends on both sets of constants, $C_{s}$ and $q_{s}$. However, one then must impose the consistency conditions \req{equalB}: all the integrals must give the same result. It turns out that these equations always give two solutions for the polarization parameters, that we denote as $q_{s}^{\pm}$ --- naturally these correspond to the two degrees of freedom of the graviton. Remarkably, this solution is always independent of the $C_{s}$ constants and the resulting values of $\delta B_{lm}$ are also independent. This can be proven analytically in the case of $m=0$ modes --- we provide an example below --- while for the rest of the modes one can check numerically that $\delta B_{lm}$ does not depend on $C_s$. This serves as another strong consistency check of our computations.  

Finally, the value of the polarization parameters $q_{s}$ that solve \req{equalB} is different in parity-preserving and in parity-breaking theories. For the former, the modes that solve \req{equalB}   are those that have a definite parity, which, as we explained in Section~\ref{sec:metricrec}, are those with 
\begin{equation} \label{eq:evenqdef}
q_{-2}^{\pm}=q_{+2}^{\pm}=\pm 1\, .
\end{equation}
On the other hand, for parity-breaking theories one cannot a priori determine the values of $q_{s}$, and in principle one must determine the  different coefficients $\delta B_{s,lm}$, $\delta B^{*}_{s,lm}$ as a function of $q_{s}$ and then solve \req{equalB}. However, a pretty remarkably observation, first noted in the appendix of \cite{Cano:2023jbk}, is that one can find a general formula for $q_{s}$ in parity-breaking theories, which turns out to be theory-independent. The result, adapted to our case, reads
\begin{equation}\label{qbreak}
q_{-2}^{\pm}=-\frac{1}{q_{+2}^{\pm}}=\frac{i D_{2}}{6m\pm \sqrt{D_{2}^2+(6m)^2}}\, ,
\end{equation}
where we recall that $D_{2}$ is given by \req{eq:D2constant}. These values appear as a solution of imposing $P_{s}=-P_{s}^{*}$ for the $P_{s}$ constants \req{eq:Pconstants}. Intuitively, the reason behind these equations is that, in an appropriate gauge, the $f_{s}$ functions that appear in the integral \req{deltaBf}, and the analogous conjugate functions $f_{s}^{*}$ that appear in the integral of $\delta B_{s,lm}^{*}$, can be shown to become proportional, and satisfy $f_{s}/f_{s}^{*}=-P_{s}^{*}/P_{s}$. Thus, imposing $\delta B_{s,lm}=\delta B_{s,lm}^{*}$ is simply equivalent to $P_{s}=-P_{s}^{*}$, which leads to \req{qbreak}. However, finding the gauge transformation that sets $f_{s}$ and $f_{s}^{*}$ to be proportional in general is challenging, so we do not have a complete proof of \req{qbreak}. Nevertheless, we have checked that in all cases \req{qbreak} leads to \req{equalB} for parity-breaking theories. Finally, a universal aspect of the theories that break parity is that the correction $\delta B_{lm}$ is opposite for each polarization, \textit{i.e.}, $\delta B^{+}_{lm}=-\delta B^{-}_{lm}$. For the theories that preserve parity one simply has independent corrections for each polarization type.

As an example, we show that $\delta B_{lm}$ does not depend on $C_{s}$ for a particular case. Let us consider, for instance, the quartic theory $\mathcal{C}^2$, with corresponding coupling constant $\epsilon_{1}$.  For simplicity, we consider the mode $l=2$ and $m=0$ and evaluate the four integrals for the correction of the angular separation constants, given by \eqref{equalB} and the conjugate ones. We find
\begin{equation}
\begin{aligned} \label{eq:evenexample}
	\delta B_{+2,lm} &=- \frac{3}{8}\alpha_{1} \left(\nu+\frac{\sigma\left(96 C_2-1\right)}{96 C_2 q_{-2} - q_{2}}\right) \,,
	\\
	\delta B_{+2,lm}^{*} &= -\frac{3}{8}\alpha_{1} \left(\nu+\frac{\sigma\left(96 C_2 q_{-2}-q_{2}\right)}{96 C_2-1}\right) \,,
	\\
	\delta B_{-2,lm} &= -\frac{3}{8}\alpha_{1}\left(\nu+\frac{\sigma \left(6 C_{-2}-1\right)}{6 C_{-2} q_{2}-q_{-2}}\right) \,,
	\\
	\delta B_{-2,lm}^{*} &=- \frac{3}{8}\alpha_{1}\left(\nu -\frac{\sigma\left(q_{-2}-6 C_{-2} q_{2}\right)}{6 C_{-2}-1}\right) \,,
\end{aligned}
\end{equation}
where $\sigma=24832+7875 \pi$, $\nu=26624+10395 \pi$, and $\alpha_{1}$ is the dimensionless coupling constant defined in \req{alphaq}. By setting these shifts equal to each other, we solve for $q_{s}$ and get precisely \eqref{eq:evenqdef}, as it should since this theory preserves parity. Imposing this back into \eqref{eq:evenexample} yields the values of $\delta B_{lm}$ for each polarization, 
\begin{align}
	\delta B^{+}_{lm} =-\alpha_{1} \frac{9}{4} (8576+3045 \pi ) \,,\quad \delta B^{-}_{lm} =-\alpha_{1} 21 (32+45 \pi )\, ,
\end{align}
which, as promised, are independent of $C_{s}$.

 \subsection{Numerical results}

Thus far, we have shown in Section~\ref{sec:generalpropertiesofthecoefficients} that the corrections to the separation constant are consistently independent of the two types of gauge freedom, $\Delta f(h(x))$ and $C_{\pm2}$. We have also discussed how two families of modes, characterized by two different polarizations $q_{s}^{\pm}$, emerge in each case. With this at hand, we proceed to compute the values of $\delta B^{\pm}_{lm}$ for various values of the angular numbers.
 
For the axisymmetric modes $m=0$, the spheroidal harmonics are analytic, for instance
  \begin{align}
 	S^{20}_{2}(x) = \frac{1}{4} \sqrt{\frac{15}{2 \pi }} \left(1-x^2\right),
 	\quad
 	S^{30}_{2}(x) = \frac{1}{4} \sqrt{\frac{105}{2 \pi }} x \left(1-x^2\right),
 \end{align}
 and more generally they are simply given by the associated Legendre polynomials, \textit{i.e.}, 
 \begin{align}
S_{s}^{l0}(x)=N_{l} P_{l}^{s}(x)\, ,
 \end{align}
 with a normalization constant $N_{l}$. 
Then, one may evaluate the integrals \req{deltaBf} analytically and find the exact value of $\delta B_{l0}$ as in the example \req{eq:evenexample} above.  In Table~\ref{table:axisymmetrictable} we show the results for the $l=2$ and $l=3$ modes of all the higher-derivative theories \req{eq:actionwithcorrections}, including odd and even parity ones. There are no corrections to the cubic odd theory for $l=2$. Moreover, as anticipated $\delta B^{\pm}_{lm}$ for the odd parity theories are equal in magnitude and opposite in sign. 
For the even-parity theories, in general one has $\delta B^{+}_{lm}\neq \delta B^{-}_{lm}$. However, for the even cubic theory we see that $\delta B^{+}_{20}=\delta B^{-}_{20}$, which connects to an observation made in \cite{Horowitz:2023xyl}. We provide the exact relation between our results and those of \cite{Horowitz:2023xyl} in Section~\ref{sec:divergence}.

  \begingroup
 \setlength{\tabcolsep}{8pt} % Default value: 6pt
 \renewcommand{\arraystretch}{1.5} % Default value: 1
 \begin{table*}[t!]
 	\centering
 	\resizebox{\textwidth}{!}{
 		\begin{tabular}{|c|c|c|c|c|}
 			\hline
 			Theory & $\displaystyle \delta B_{20}^{+}/\alpha_q$ & $\displaystyle\delta B_{20}^{-}/\alpha_q$ & $\displaystyle\delta B_{30}^{+}/\alpha_q$ & $\displaystyle \delta B_{30}^{-}/\alpha_q$ %& $\delta B_{40}^{+}/\alpha_q$ & $\delta B_{40}^{-}/\alpha_q$
 			\\ \hline \hline
 			$\lambda_{\rm ev}$ & $\displaystyle\frac{120}{7}$ & $\displaystyle\frac{120}{7}$ & $\displaystyle 360 (-551+175 \pi )$ & $\displaystyle 2520 (63-20 \pi)$ %& $\frac{5400}{7} (15435 \pi -48491)$ & $-\frac{3240}{7} (23520 \pi -73891)$  
 			\\ \hline 
 			$\lambda_{\rm odd}$ & $\displaystyle 0$ & $\displaystyle 0$ & $\displaystyle 180 (992-315 \pi)$ & $\displaystyle 180 (-992+315 \pi )$ %& $-3780 (3015 \pi -9472)$ & $3780 (3015 \pi -9472)$ 
 			\\ \hline
 			$\epsilon_{1}$ &  $\displaystyle -\frac{9}{4} (8576+3045 \pi )$ & $\displaystyle -21 (32+45 \pi )$ & $\displaystyle \frac{21}{2} (1216-1575 \pi)$ & $\displaystyle \frac{1323}{4} (384-175 \pi)$
 			\\ \hline
 			$\epsilon_{2}$ &  $\displaystyle -\frac{189}{4} (384+145 \pi )$ & $\displaystyle -2208-945 \pi$  & $\displaystyle \frac{21}{2} (704-1575 \pi )$ & $\displaystyle \frac{63}{4} (8576-3675 \pi)$ %& $\frac{567}{4} (588416-187065 \pi )$ & $\frac{88154325 \pi }{4}-69117408$ 
 			\\ \hline
 			$\epsilon_{3}$ & $\displaystyle 672$ & $\displaystyle -672$ & $\displaystyle -3360$ & $\displaystyle 3360$ %& $-756 (43785 \pi -137528)$ & $756 (43785 \pi -137528)$ 
 			\\ \hline
 	\end{tabular}}
 	\caption{The analytical values of the corrections to the separation constants for the axisymmetric modes $m=0$ and $l=2,3$. We normalize $\delta B_{lm}$ by the corresponding dimensionless coupling $\alpha_{\rm q}$, defined in \req{alphaq}.} 
 	\label{table:axisymmetrictable}
 \end{table*}
 \endgroup 
 
 \begingroup
 \setlength{\tabcolsep}{6pt} % Default value: 6pt
 \renewcommand{\arraystretch}{1.5} % Default value: 1
 \begin{table*}[t!]
 	\centering
 	\resizebox{\textwidth}{!}{
 		\begin{tabular}{|c|c|c|c|c|c|c|}
 			\hline
 			Theory & $\delta B_{20}^{+}/\alpha_q$ & $\delta B_{20}^{-}/\alpha_q$ & $\delta B_{21}^{+}/\alpha_q$ & $\delta B_{21}^{-}/\alpha_q$ & $\delta B_{22}^{+}/\alpha_q$ & $\delta B_{22}^{-}/\alpha_q$
 			\\ \hline \hline
 			$\lambda_{\rm ev}$ & $1.714\times 10^1$ & $1.714\times 10^1$ & $5.080$ & $-2.069\times 10^1$ & $2.775\times 10^1$ & $3.833\times 10^1$ 
 			\\ \hline
 			$\lambda_{\rm odd}$ &  $0$ & $0$ & $-1.371\times 10^1$ & $1.371\times 10^1$ & $1.228\times 10^1$ & $-1.228\times 10^1$ 
 			\\ \hline
 			$\epsilon_{1}$  & $-4.082\times 10^4$ & $-3.641\times 10^3$ & $-6.259\times 10^3$ & $-2.465\times 10^4$ & $-8.356\times 10^3$ & $-6.285\times 10^3$ 
 			\\ \hline
 			$\epsilon_{2}$ & $-3.967\times 10^4$ & $-5.177\times 10^3$ & $-6.046\times 10^3$ & $-2.444\times 10^4$ & $-8.322\times 10^3$ & $-6.282\times 10^3$ 
 			\\ \hline
 			$\epsilon_{3}$ & $6.720\times 10^2$ & $-6.720\times 10^2$ & $-4.167\times 10^{-1}$ & $4.167\times 10^{-1}$ & $1.746\times 10^1$ & $-1.746\times 10^1$
 			\\ \hline
 	\end{tabular}}
 	\caption{The numerical corrections to the separation constant for $l=2$ and $m=0,1,2$. We normalize $\delta B_{lm}$ by the corresponding dimensionless coupling $\alpha_{\rm q}$, defined in \req{alphaq}.}
 	\label{table:numericaltable} 
 \end{table*}
 \endgroup
Departing from $m=0$ requires numerical integration of \req{deltaBf}, since the spin-weighted spheroidal harmonics are not known analytically.  These functions can be found numerically for a wide range of values of $l$ and $m$ using the package ``\textit{SpinWeightedSpheroidalHarmonics}'' of the Black Hole Perturbation Toolkit \cite{BHPToolkit}. The numerical integration must be done carefully as the integrand becomes highly oscillatory for higher $l$ and $m$, and we need to increase the precision of the numerical method as we increase the angular numbers. To test our results, we compute the four integrals corresponding to $\delta B_{s,lm}$ and $\delta B_{s,lm}^{*}$ for the appropriate values of the polarization (either \req{eq:evenqdef} or \req{qbreak}, depending on the parity of the theory) and for several values of $C_{s}$. We obtain within numerical precision that the four integrals agree and that they are independent of $C_{s}$, hence providing a very strong consistency check of our results. 
   \begin{figure}[H]
  	\centering
  	\includegraphics[width=0.99\textwidth]{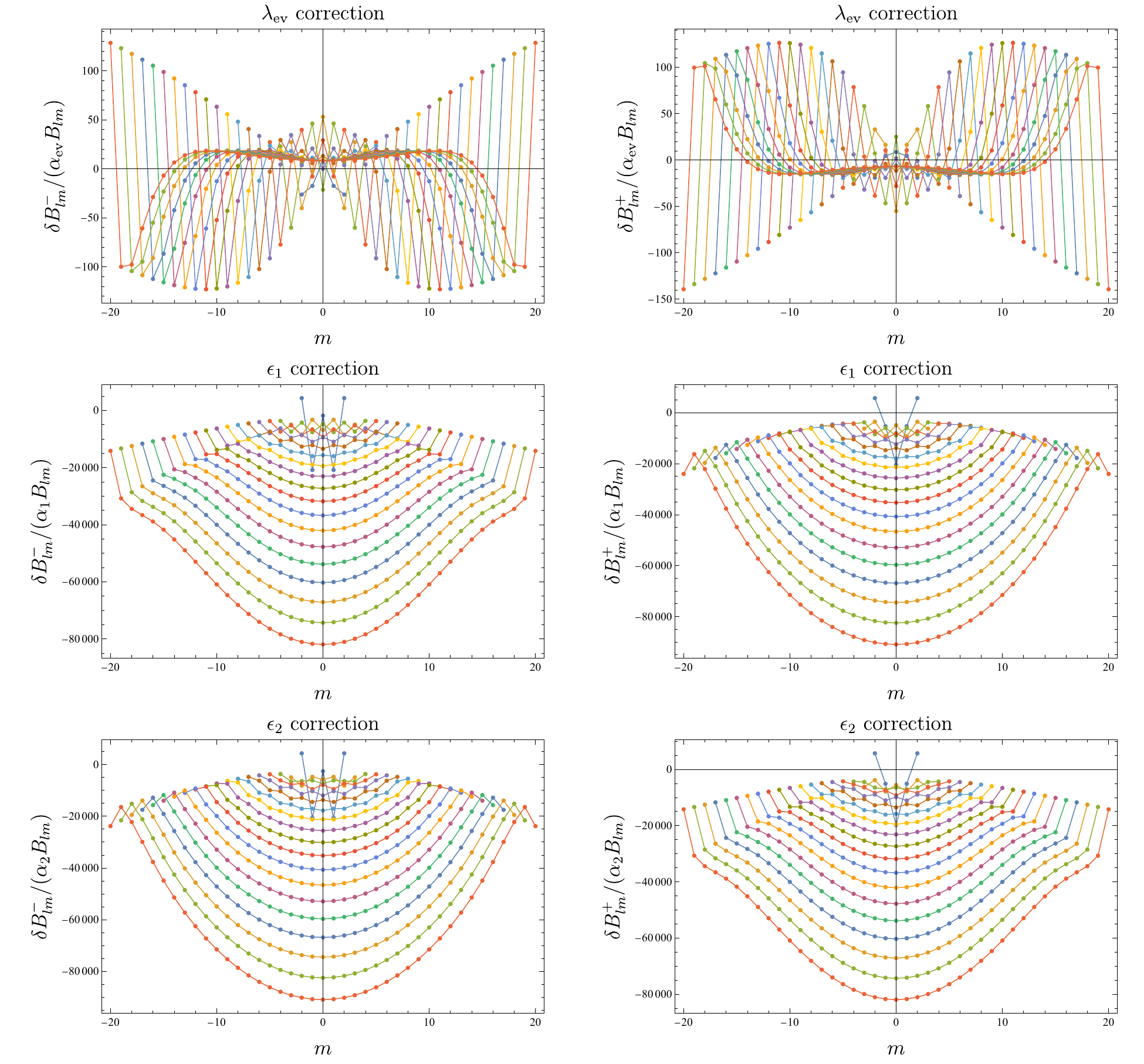}
	\includegraphics[width=0.99\textwidth]{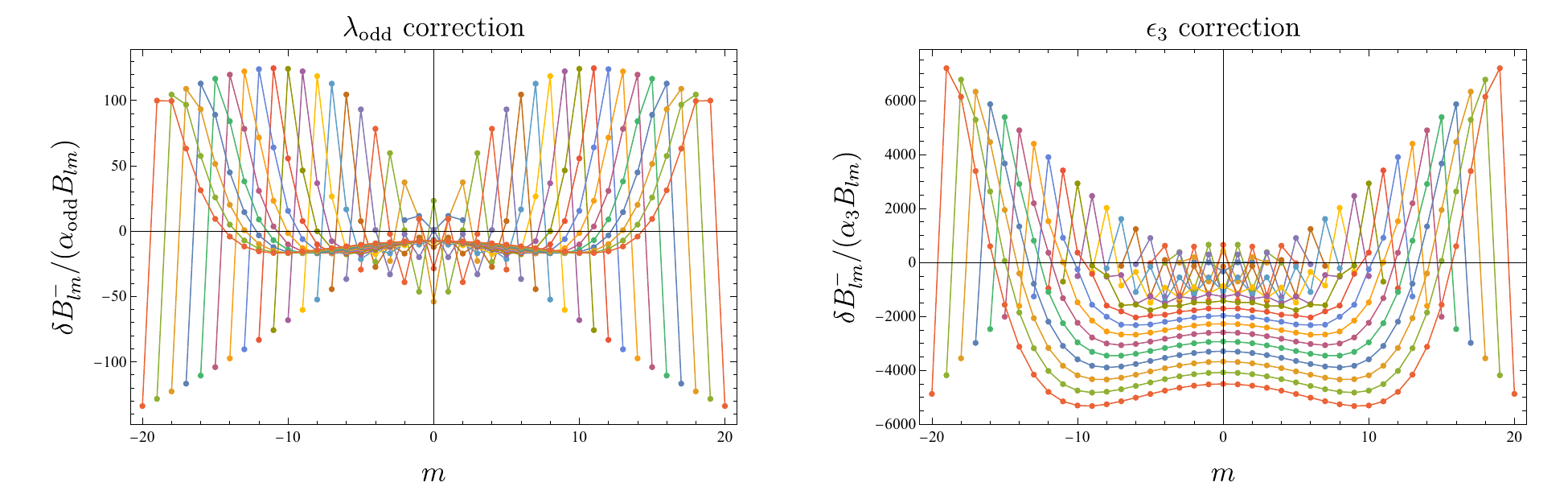}
  	\caption{The relative values of the corrections to the angular separation constants for $l=2$ to $l=20$ for all the theories in the effective action \req{eq:actionwithcorrections}. The lines connect points with the same $l$ and different $m$. Since the odd-parity theories have equal-in-magnitude corrections for both polarizations, we choose to only show the plot for the negative polarization.}
  	\label{fig:evenplots}
  \end{figure}
We show a few of the $\delta B_{lm}$ values in Table~\ref{table:numericaltable} for both polarizations, and more conclusive plots for the relative change $\delta B^{\pm}_{lm}/(\alpha_{q} B_{lm})$ given for $l=2$ to $l=20$ in Figure~\ref{fig:evenplots}. In these plots, the lines connect points with the same $l$ and $m=-l,\dots, l$. 

The first thing we notice is that these coefficients are symmetric under $m\rightarrow -m$, so $\delta B_{l-m}=\delta B_{lm}$. 
Furthermore, we can see that the corrections take a particular pattern for each theory. For example, in both cubic theories, and for both polarizations, the coefficients $\delta B_{lm}$ have a similar structure and the magnitude of the corrections seems to begin increasing rapidly once the value of $|m|$ approaches $l$. On the other hand, the even-parity quartic theories have a well-shaped pattern for low values of $|m|$ relative to $l$ and then distinctively changes when $|m| \sim l$, producing a butterfly-like shape. 
On the other hand, from these plots it is more or less evident that the relative correction $\delta B_{lm}/B_{lm}$ grows with $l$ at a higher rate for the quartic theories than for the cubic theories. 
However, for large values of $l$, the numerical integration becomes more difficult as there is increasing error in the results of $\delta B_{lm}$, so these modes cannot be accessed numerically. Fortunately, this regime can be studied analytically, as we explore in the next section.

 \subsection{Eikonal limit}\label{sec:eikonal}
Eikonal modes are those in the limit of large momentum $l\rightarrow \infty$, $m\rightarrow \infty$, with $l/m$ fixed. The eikonal limit is interesting due to its connection with wave propagation and its geometric interpretation \cite{Cardoso:2008bp,Yang:2012he}. 
The expressions of $f_s$ in \req{deltaBf} greatly simplify in this limit. For the cubic theories we have\footnote{In the case of $f_{+2}^{(\rm odd)}$ we have removed a term that is independent of $q_2$ and that vanishes upon performing the integration \req{deltaBf}.}
\begin{align}\notag
f_{+2}^{(\rm ev)}=&-\alpha_{\rm ev}\Bigg[\frac{180 (2-i x) \left(4 B_{lm} \left(x^2-1\right)+m^2 \left(2 x^4+x^2+5\right)\right)}{q_{2} (x-i)^8}\\\notag
&-\frac{1}{14 \left(x^2+1\right)^6} \Big(16 B_{lm} (131 x^{10}+414 x^8-630 i x^7-2898 x^6+6678 i x^5+12984 x^4\\\notag
&-9954 i x^3-12289 x^2+2898 i x+986)+m^2 (107 x^{14}+843 x^{12}+1311 x^{10}-2016 i x^9\\\notag
&-10285 x^8+23688 i x^7+52961 x^6-51912 i x^5-99559 x^4+119448 i x^3+174037 x^2\\\label{feveikonal}
&-44856 i x-15863)\Big)\Bigg]\, ,\\
f_{+2}^{(\rm odd)}=&\alpha_{\rm odd}\frac{180 i (2-i x) \left(4 B_{lm} \left(x^2-1\right)+m^2 \left(2 x^4+x^2+5\right)\right)}{q_{2} (x-i)^8}\, ,
\end{align}
while for the quartic theories 
\begin{align}\label{feps1}
f_{+2}^{(1)}&=36\alpha_{\rm 1}\frac{\left(4 B_{lm}-3 m^2\right)^2 \left((x+i)^5-q_{2} (x-i)^5\right)}{q_{2} (x-i)^9 (x+i)^4}\, ,\\\label{feps2}
f_{+2}^{(2)}&=36\alpha_{\rm 2}\frac{\left(4 B_{lm}-3 m^2\right)^2 \left(-(x+i)^5-q_{2} (x-i)^5\right)}{q_{2} (x-i)^9 (x+i)^4}\, ,\\
f_{+2}^{(3)}&=-36\alpha_{\rm 3}\frac{\left(4 B_{lm}-3 m^2\right)^2 i (x+i) }{q_{2} (x-i)^9}\, ,
\label{f3eikonal}
\end{align}
where we recall that the $\alpha_{\rm q}$ constants are defined in \req{alphaq}, and where we are neglecting subleading terms in the large $l$ and $m$ expansion. Also, we note that 
\begin{equation}
B_{lm}=l^2\left[1-\frac{1}{8}\frac{m^2}{l^2}\left(1-\frac{m^2}{l^2}\right)\right]
\end{equation}
at leading order in the eikonal expansion \cite{Yang:2013uba}.
From these expressions, it already follows that the scaling of $\delta B_{lm}$ goes as $l^2$ in the case of the cubic theories, and as $l^4$ for the quartic theories. However, this result is still unable to capture the full dependence on $l$ and $m$, since one still has to compute the integral \req{deltaBf}. In the eikonal limit, this integral is highly oscillatory and thus its numerical evaluation is challenging. 

Remarkably, there is a way of computing \req{deltaBf} analytically in the eikonal limit. To this end, we recall the gauge ambiguity in the integrand that we studied above, which allows us to consider the family of integrals
  \begin{equation}\label{deltaBf2}
 \delta B_{lm}=2\pi \int_{-1}^{1}dx  \, \left(S_{s}^{lm}(x)\right)^2 \left[f_{s}(x)+\Delta f_{s}(h(x))\right]\, ,
 \end{equation}
 where $\Delta f_{s}$ is given by \req{eq:gaugeterm} and $h(x)$ is an arbitrary smooth function that vanishes at $x=\pm1$. Since the result is independent of $h(x)$, the idea is to choose a function that simplifies the integrand. In particular, if we can find an $h$ such that 
 \begin{equation}\label{kappaeq}
f _{s}(x)+\Delta f_{s}(h(x))=\kappa\, ,
\end{equation}
for a constant $\kappa$, then the integral would be trivial due to the normalization of $S_{s}^{lm}$ \req{ortho}, and we would simply have $\delta B_{lm}=\kappa$. Of course, this means that the value of $\kappa$ cannot be arbitrary: it is determined by the condition that the equation \req{kappaeq} has a regular solution. The analysis of the regularity of the solutions of \req{kappaeq} is challenging in the general case, since it is a third-order differential equation. However, in the eikonal limit, the operator $\Delta f_{s}(h(x))$ becomes
 \begin{equation}
\Delta f_{s}(h)= -\left(2 B_{lm}-\frac{m^2 \left(x^4-x^2+4\right)}{2 \left(1-x^2\right)}\right) h'(x)+\frac{x \left(4 B_{lm} \left(x^2-1\right)+m^2 \left(x^2+7\right)\right)}{2 \left(1-x^2\right)^2}h(x)\, ,
\end{equation}
plus subleading terms in the eikonal expansion that we neglect. Thus \req{kappaeq} is now a first-order differential equation that we can easily integrate, and its solution reads
 \begin{figure}[t!]
	\centering
	\includegraphics[width=0.99\textwidth]{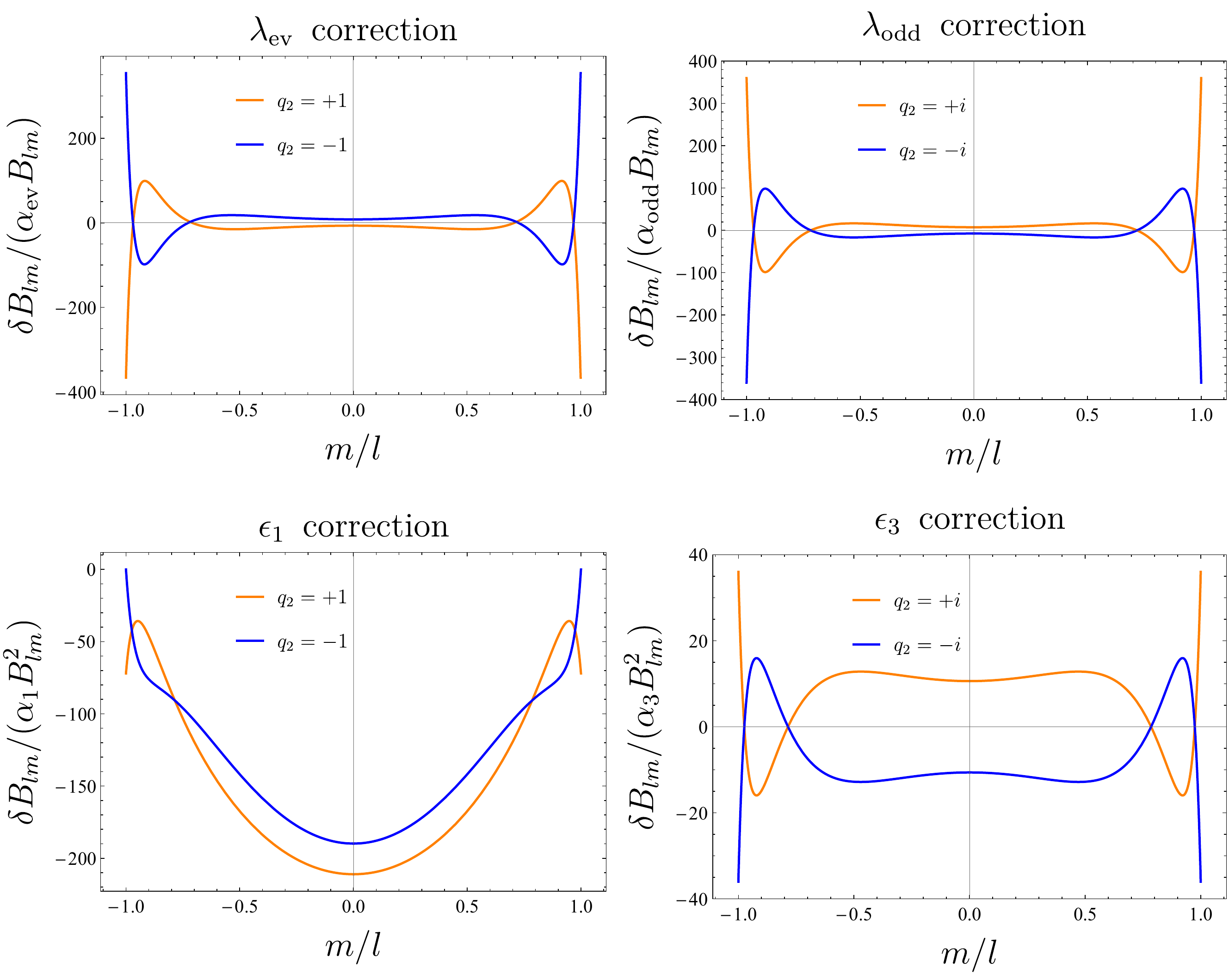}
	\caption{Corrections to the angular separation constants in the eikonal limit as a function of $m/l$.  We normalized $\delta B_{lm}$ by the corresponding dimensionless coupling constant and by the appropriate power of $B_{lm}\sim l^2$. We do not show the case of the quartic $\epsilon_2$ theory since it is equivalent to the $\epsilon_{1}$ theory upon exchanging the polarization. }
	\label{fig:eikonal}
\end{figure}

\begingroup
\setlength{\tabcolsep}{8pt} % Default value: 6pt
\renewcommand{\arraystretch}{1.5} % Default value: 1
\begin{table*}[t!]
	\centering
	\begin{tabular}{|c|c|c|c|c|}
	\hline
Theory & $ \delta B_{l0}^{+}/\alpha_{\rm q}$ & $\delta B_{l0}^{-}/\alpha_{\rm q}$ & $\delta B_{ll}^{+}/\alpha_{\rm q}$ & $\delta B_{ll}^{-}/\alpha_{\rm q}$\\
 \hline \hline
$\lambda_{\rm ev}$&$\displaystyle -\frac{615 l^2}{64 \sqrt{2}}$ &$\displaystyle \frac{735 l^2}{64 \sqrt{2}}$ & $\displaystyle-\frac{5127
   l^2}{14}$ & $\displaystyle\frac{4953 l^2}{14}$ \\\hline
 $\lambda_{\rm odd}$&$\displaystyle\frac{675 l^2}{64 \sqrt{2}}$ & $\displaystyle-\frac{675 l^2}{64 \sqrt{2}}$ & $\displaystyle 360 l^2$ &
   $\displaystyle-360 l^2$ \\\hline
$\epsilon_{\rm 1}$& $\displaystyle-\frac{76419 l^4}{256 \sqrt{2}}$ & $\displaystyle-\frac{68733 l^4}{256 \sqrt{2}}$ & $\displaystyle-72
   l^4 $& $\displaystyle 0$ \\\hline
$\epsilon_{\rm 2}$& $\displaystyle-\frac{68733 l^4}{256 \sqrt{2}}$ & $\displaystyle-\frac{76419 l^4}{256 \sqrt{2}}$ & $ \displaystyle 0$ &
   $\displaystyle-72 l^4$ \\\hline
$\epsilon_{\rm 3}$& $\displaystyle\frac{3843 l^4}{256 \sqrt{2}}$ & $\displaystyle-\frac{3843 l^4}{256 \sqrt{2}}$ & $\displaystyle 36 l^4$ &
   $\displaystyle -36 l^4$ \\
   \hline
	\end{tabular}
	\caption{Correction to the angular separation constants in the eikonal limit for $m=0$ and $m=l$, for each theory, and for both possible polarizations. We show $\delta B_{lm}$ normalized by the corresponding dimensionless coupling $\alpha_{\rm q}$, defined in \req{alphaq}.}
	\label{table}
\end{table*}
\endgroup

\begin{equation}\label{eq:hsol}
h(x)=-\frac{(1-x^2)}{\sqrt{Z(x)}}\left[c+2\int dx\frac{f_{s}(x)-\kappa}{\sqrt{Z(x)}}\right]\, ,
\end{equation}
where $c$ is an integration constant and
\begin{equation}
Z(x)=4B_{lm}\left(1-x^2\right)-m^2 \left(x^4-x^2+4\right)\, .
\end{equation}
 We check that $Z(0)=4(B_{lm}-m^2)>0$ for $m\in [-l,l]$ for large enough $l$, and thus it is always positive in the eikonal limit. On the other hand, $Z(\pm 1)=-m^2\le 0$. Therefore, $Z(x)$ always has two roots $Z(\pm x_0)=0$ for $x_0\in (0,1]$.  The position of the root $x_0$ can be used to parametrize the ratio $m^2/B_{lm}$, which goes between $0$ and $1$,
 \begin{equation}
 \frac{m^2}{B_{lm}}=\frac{4(1-x_0^2)}{4+x_0^2-x_0^4}\, .
 \end{equation}
 Thus, the solution $h(x)$ is in general singular at $x=\pm x_0$. However, there is a unique value of $\kappa$ and of the integration constant $c$ such that the solution becomes fully regular. One can see that the condition on $\kappa$ is
 \begin{equation}\label{eq:hcond}
 \int_{-x_0}^{x_0} dx\frac{f_{s}(x)-\kappa}{\sqrt{Z(x)}}=0\, .%\quad \Rightarrow\quad \kappa= \frac{\int_{-x_0}^{x_0} dx Z(x)^{-1/2}f_{s}(x)}{\int_{-x_0}^{x_0} dx Z(x)^{-1/2}}\, .
 \end{equation}
 Furthermore, once we have removed the singularities of $h(x)$ at $x=\pm x_0$, we observe from \req{eq:hsol} that $h(\pm1)=0$, which guarantees that the integral \req{deltaBf} is in fact unaffected by $h(x)$.  

From \req{eq:hcond} we can solve for $\kappa$ straighforwardly, and making the change of variable $x=x_0 y$, and taking into account that $\delta B_{lm}=\kappa$, we arrive at the result 
 
 \begin{equation}\label{magicformula}
 \delta B_{lm}=\frac{1}{2K\left(-k\right)}\int_{-1}^{1} dy\frac{f_s(x_0 y)}{\sqrt{(1-y^2)(1+k y^2)}}\,, %\quad \text{where}\,\, k=\frac{x_0^2(-1+x_0^2)}{3+x_0^2}\, ,
 \end{equation} 
 where 
 \begin{equation}\label{kdef}
 k=\frac{x_0^2(1-x_0^2)}{3+x_0^2}\, ,
 \end{equation} 
 and
 \begin{equation}\label{ellipk}
 K(-k)=\int_{0}^{1} dy\frac{1}{\sqrt{(1-y^2)(1+k y^2)}}
 \end{equation}
 is the complete elliptic integral of the first kind.  We can now plug the functions \req{feveikonal}-\req{f3eikonal} in \req{magicformula} and perform the integration, which can be expressed analytically in terms of elliptic integrals. Since the result is somewhat convoluted, we show the full expressions in the Appendix \ref{app:eikonal}.

 As we anticipated, the correction $\delta B_{lm}$ scales as $l^2$ in the case of the cubic theories and as $l^4$ in the case of quartic theories. It is actually pretty remarkable that in the cubic case $\delta B_{lm}$ scales as fast as $B_{lm}\sim l^2$, which implies that the cubic EFT does not break down in the eikonal limit as long as the coupling constant is small enough. On the other hand, since for the quartic corrections $\delta B_{lm}$ grows faster than $B_{lm}$, the corrections always become dominant for large enough $l$ and thus the EFT breaks down in the eikonal limit.  This generically happens for $l\sim |\alpha_{q}|^{-1/2}$, up to a numerical factor.  
 On the other hand, an interesting property of the even-parity quartic corrections, which follows from \req{feps1} and \req{feps2}, is that they yield the same values of $\delta B_{lm}$ but swapping the polarizations, that is $\delta B_{lm}^{\pm }(\epsilon_1)=\delta B_{lm}^{\mp }(\epsilon_2)$. 
 
 In Figure \ref{fig:eikonal} we show the profile of $\delta B_{lm}$ (normalized by its scaling $l^n$ and by the corresponding coupling constant $\alpha_{\rm q}$) as a function of the ratio $m/l$.  These plots are obtained from the expressions of $\delta B_{lm}$ in Appendix \ref{app:eikonal}. We omit the quartic $\epsilon_2$ theory since the curves are the same as for the $\epsilon_1$ theory, on account of the symmetry we just mentioned above. As we can see, these curves resemble very closely the profile of $\delta B_{lm}$ in Figure \ref{fig:evenplots} that we obtained numerically for finite values of $l$. 
Finally, in Table~\ref{table} we show the analytic results of $\delta B_{lm}$ for the cases of $m=0$ and $m=l$.  We observe that $\delta B_{ll}^{+}$ and $\delta B_{ll}^{-}$ vanish, respectively, for the $\epsilon_1$ and $\epsilon_2$ corrections. However, this only means that they vanish to leading order in the eikonal expansion, and thus we expect that they are actually of order $\mathcal{O}(l^3)$ instead of $\mathcal{O}(l^4)$. 
 
 \section{Divergence of tidal forces}\label{sec:divergence}
We now turn our attention to the corrected Teukolsky radial equation \req{eq:RadialCorrected}.  Although, as we observed, the only effect of the corrections is to modify the angular separation constants, this has interesting implications.
Here we reproduce --- and extend --- the results of \cite{Horowitz:2023xyl}, which showed that the perturbed near-horizon geometries in higher-derivative gravity generically develop singularities. 
 
 Let us consider, as in \cite{Horowitz:2023xyl}, the case of static perturbations, $\om=0$. The Teukolsky radial equation \req{eq:RadialCorrected} simplifies to 
 \begin{equation}
\begin{aligned}
\rho^{-2s}\frac{d}{d\rho}\left[\rho^{2s+2}\frac{dR_{s}}{d\rho}\right]+\left(s+\frac{7m^2}{4}-B_{lm}-\delta B_{lm}\right)R_{s}&=0\, ,\\
\end{aligned}
\end{equation}
whose physically relevant solution reads
\begin{equation}
R_{s}(\rho)=A  \rho^{\gamma_{slm}}\, ,
\end{equation}
where $A$ is an integration constant and the exponent $\gamma_{slm}$ is given by
\begin{equation}\label{gammaslmdef}
\gamma_{slm}=\frac{1}{2}\left(-1-2s+\sqrt{17+4B_{lm}+4\delta B_{lm}-7m^2}\right)\, .
\end{equation}
There is another solution with an exponent that has a minus sign in front of the square root, but that one is always pathological, so we discard it. 
Let us further consider, as in \cite{Horowitz:2023xyl}, the case of axisymmetric perturbations, $m=0$. Then, expanding to first order in the corrections $\delta B_{lm}$, and using that $B_{l0}=l(l+1)-4$, the exponent takes the simple form
\begin{equation}
\gamma_{sl0}=l-s+\frac{\delta B_{l0}}{1+2l}\, .
\end{equation}
Thus, we see that, in the case of Einstein gravity, $\gamma_{sl0}$ is always a non-negative integer, $\gamma_{sl0}\in \{0\}\cup \mathbb{N}$. Therefore, the Teukolsky variables are simply polynomials in $\rho$ and they are smooth. On the other hand, once we take into account the corrections, these exponents are no longer integers so that the perturbations are no longer of class $\mathcal{C}^{\infty}$. This is particularly important for the mode $s=l=2$, since in that case we have
\begin{equation}
\gamma_{220}=\frac{\delta B_{20}}{5}\, .
\end{equation}
Therefore, if $\delta B_{20}<0$ (which depends on the sign of the higher-derivative couplings), the perturbed Weyl scalar $\Psi_{0}\sim \rho^{\gamma_{220}}$  diverges on the horizon, indicating a divergence of tidal forces. By using the values of $\delta B_{20}^{\pm}$ for both polarizations in Table~\ref{table:axisymmetrictable}, we check that our values of $\gamma_{220}$ perfectly match those of \cite{Horowitz:2023xyl}, computed via metric perturbations.\footnote{In the case of $\epsilon_2$ corrections --- denoted there by the coupling $\tilde \lambda$ --- there is factor 4 of difference due to the factor of $1/2$ that we introduce in the definition of the dual Riemann tensor \req{dualRiem}.}  We note that the polarization that we denote here by ``$+$''  corresponds to the family of modes labeled as ``$-$'' in \cite{Horowitz:2023xyl}, and viceversa. 
As discussed in \cite{Horowitz:2023xyl}, in the case of the even parity quartic theories,  $\Psi_0$ would diverge for positive couplings $\epsilon_{1}>0$,  $\epsilon_{2}>0$ (as this leads to $\gamma_{220}<0$), which is the sign selected by causality \cite{Gruzinov:2006ie}. For the even parity cubic theory, the divergence takes place if $\lambda_{\rm ev}<0$ instead. 

However, we find that \cite{Horowitz:2023xyl} seems to have missed that parity-breaking corrections do contribute to $\gamma_{220}$. While for the odd parity cubic theory we do find  $\gamma_{220}=0$, this seems to be an accident. In fact, for the quartic theory $\epsilon_{3}$, we have 
\begin{equation}\label{gammaquartic3}
\gamma_{220}=\pm \frac{672 \epsilon_{3}\ell^6}{5\mu^6}\, ,
\end{equation}
where each sign corresponds to a different polarization. Therefore, not only the parity-breaking $\epsilon_{3}$ theory affects this result, but we find something remarkable: $\Psi_{0}$ diverges for any sign of $\epsilon_3$, since there is always one polarization for which $\gamma_{220}$ is negative. 

Thus, the conclusion, as in \cite{Horowitz:2023xyl}, is that some modes that used to be regular in Einstein gravity develop curvature singularities at the horizon in the case of higher-derivative gravity. 

We would like to point out that there are other modes that are already singular in Einstein gravity.  In fact, there are non-axisymmetric modes for which the square root \req{gammaslmdef} becomes imaginary and they diverge strongly at $\rho=0$ even in the absence of corrections. This is briefly mentioned in \cite{Horowitz:2022mly}, which points out that these modes must be related to superradiance. In fact, we recall that perturbations with $\om=0$ and $m\neq 0$ at the throat of extremal Kerr, actually correspond to modes at the critical frequency for superradiance $\omega=m\Omega_{H} $ at infinity. However, in our opinion, the status of these singular non-axisymmetric modes is pretty unclear and it would be interesting to explore their significance elsewhere. See \cite{Casals:2016mel} though for a discussion of similar modes in the case of perturbations of a scalar field on a extremal Kerr background. 

The singularity found for $m=0$ modes in higher-derivative gravity is of a different nature (unrelated to superradiance), since these modes are actually static.  This divergence implies that the black hole geometry becomes singular when it is deformed by an axisymmetric external field. It would be interesting to investigate the phenomenological consequences of this divergence by studying the full black hole solutions beyond the near-horizon region.

 %%%%%%%%%%%%%%%%%%%%%%%%%%%%%%%%%%%%%%%%%%%%%%%%%%%%%%%%%%%%%%
%%%%%%%%%%%%%%%%%%%%%%%%%%%%%%%%%%%%%%%%%%%%%%%%%%%%%%%%%%%%%%
\section{Connection to the full radial Teukolsky equation} \label{sec:connectionstofullTequation}
%%%%%%%%%%%%%%%%%%%%%%%%%%%%%%%%%%%%%%%%%%%%%%%%%%%%%%%%%%%%%%
%%%%%%%%%%%%%%%%%%%%%%%%%%%%%%%%%%%%%%%%%%%%%%%%%%%%%%%%%%%%%%
Having derived the (near)-extremal near-horizon limit of the Teukolsky equation for higher-derivative theories, now our goal is find what information it can give us about the full Teukolsky radial equation in those theories. 
It has been shown in \cite{Cano:2023jbk} that the full radial Teukolsky equation for any sub-extremal black hole in higher-derivative gravity can always be written as
\begin{align} \label{eq:Twithcorrections}
	\Delta^{-s+1}\frac{d}{dr}\left[\Delta^{s+1}\frac{dR_s}{dr}\right] + \left(V_{s}+\alpha \delta V_{s}\right)R_{s} = 0 \,,
\end{align}
where the potential $V_s$ [in Eq. \req{TPoten}] is corrected by $\delta V_{s}$, and $\Delta$ is given by the original Kerr function \req{Delta}.
Thus, our idea is to take the near-extremal near-horizon limit of \req{eq:Twithcorrections} and to compare the result with \req{Tnext} in order to set constraints on $\delta V_{s}$ near extremality. 

However, \req{eq:Twithcorrections} is problematic at extremality, because in higher-derivative gravity the extremality condition is modified and it takes place for
\begin{equation}\label{eq:extbound}
a_{*}=M(1+\alpha \xi_{1})\, ,
\end{equation}
for some constant $\xi_{1}$.  On the other hand, $\Delta$ has a double root for $a=M$, not for $a=a_{*}$. Thus, this form of the equation is not appropriate to take the (near-)extremal limit. In order to remedy this, we perform the following change of variables to first order in $\alpha$
\begin{equation}
\hat{R}_{s}=R_{s}-\alpha\frac{2\xi_{1}M^2}{r_{+}-r_{-}}\left(\frac{(s+1)}{r-r_{-}}R_{s}+\frac{dR_{s}}{dr}\right)\, ,
\end{equation}
where $r_{\pm}$ are those in \req{rpmKerr}. This transformation puts the equation into the form 
\begin{align} \label{eq:Twithcorrections2}
	\hat\Delta^{-s+1}\frac{d}{dr}\left[\hat\Delta^{s+1}\frac{d\hat{R}_s}{dr}\right] + \left(V_{s}+\alpha \delta \hat{V}_{s}\right)\hat{R}_{s} = 0 \,,
\end{align}
where 
\begin{equation}
\hat \Delta =r^2-2M r+a^2+M^2-a_{*}^2\, ,
\end{equation}
and 
\begin{align}\label{dVtransf}
	\delta \hat{V}_{s}&=\delta V_{s}+\frac{2\alpha  \xi _1  M^2}{r_{+}-r_{-}} \left(V_s'-\frac{4 V_s}{r-r_-}+ \left(r-r_+\right) (s+1)^2+\frac{3 \left(r-r_+\right){}^2}{\left(r-r_-\right)}\right)+\mathcal{O}(\alpha^2)\, .
\end{align}
Observe that now $\hat \Delta$ develops a double root at extremality $a=a_{*}$, since we have the roots
\begin{equation}
\hat{r}_{\pm}=M\pm \sqrt{a_{*}^2-a^2}\, .
\end{equation}
This comes at the price of introducing a term that diverges in the extremal limit (when $r_{+}\sim r_{-}$) in the potential. The idea is that the original potential $\delta V_{s}$ is in fact singular in the extremal limit, while $\delta \hat{V}_{s}$ is fully regular, so the term introduced by the change of variables must precisely cancel the divergences of $\delta V_{s}$.

A caveat in this computation is that we are performing an expansion at first order in $\alpha$, and this is only valid  if we are far enough from extremality. For instance, if we expand $\hat{r}_{+}$ we obtain 
\begin{equation}
\hat{r}_{+}=M+\sqrt{M^2-a^2}+\frac{\alpha  \xi_{1} M^2}{\sqrt{M^2-a^2}}+\mathcal{O}(\alpha^2)\, ,
\end{equation}
which contains a term that apparently diverges when $a\rightarrow M$. However, this only indicates that the series expansion at first order in $\alpha$ breaks down when $M\sim a$. Thus, we restrict ourselves to the ``far near-extremal'' regime
\begin{equation}
1\gg\left|1-\frac{a^2}{M^2}\right|\gg \alpha\, ,
\end{equation}
where the linear expansion in $\alpha$ is valid.  Observe that this regime means that the temperature of the black hole, although much smaller than $1/M$, is still much larger than $\alpha^{1/2}$, $T\gg \alpha^{1/2}/M$.
On the other hand, the ``ultra near-extremal'' regime would correspond to 
\begin{equation}
\left|1-\frac{a^2}{M^2}\right|\sim \alpha\ll 1\, .
\end{equation}
In this case $T\sim \alpha^{1/2}/M$, which is the reason why the linear $\alpha$ expansion breaks down --- see \textit{e.g.} \cite{Goon:2019faz,Cano:2019ycn} for discussions of the same phenomenon in the case of charged black holes with higher-derivative corrections.  Thus, this case requires a different analysis that may be carried out elsewhere.

Let us then study the near-extremal near-horizon limit of \req{eq:Twithcorrections2}. The limit must be taken in the following form 
\begin{align}\label{newrules}
		r &=r_{+} + \epsilon \rho \left(1+\frac{\alpha}{2}(\xi_{1}+\xi_{2}-2\xi_{4})\right), \quad t =2M^2(1+\alpha \xi_{4}) \frac{\tau}{\epsilon}, \quad \phi= \psi +  \Omega_{\rm H} t,\\
	 a&=a_{*} - \frac{\rho_0^2}{8M}(1+\alpha (\xi_{2}-2\xi_{4}))\epsilon^2\,,\quad \omega=m \Omega_{\rm H}+\frac{\epsilon\, \om}{2M^2}(1-\alpha\xi_{4})\, . 
	 \label{newrules2}
\end{align}
and where $\Omega_{\rm H}$ is the angular velocity of the horizon,
\begin{equation}
\Omega_{\rm H}=\frac{1}{2M}\left(1+\alpha\xi_{3}\right)\, .
\end{equation}
These expressions contain several coefficients $\xi_{i}$ that appear because the higher-derivative corrections modify the way of taking the near-horizon limit. As we have seen,  $\xi_{1}$ measures the correction to the extremality condition \req{eq:extbound}, while  $\xi_{3}$ is nothing but the correction to the angular velocity of the horizon.  On the other hand, the coefficient $\xi_{4}$ modifies the possible relation between the asymptotic time $t$ and the near-horizon time $\tau$, which is chosen by the condition that the AdS$_2$ metric takes the canonical form $-\rho(\rho+\rho_0)d\tau^2+(\rho(\rho+\rho_0))^{-1}d\rho^2$.  Finally, the parameter $\xi_{2}$ appears in the relation between temperature and angular momentum. Indeed, note that $\rho_0$ always corresponds to the near-horizon temperature by $\rho_0=4\pi T_{\rm nh}$, while the temperature measured at infinity is obtained by applying the rescaling between $t$ and $\tau$ in \req{newrules}, so we obtain
\begin{equation}
T=\frac{\epsilon \rho_0}{8\pi M^2(1+\alpha \xi_{4})}\, .
\end{equation}
Therefore, the relation for $a$ in \req{newrules2} is equivalent to the small-temperature expansion of the relation $a(M,T)$, 
\begin{equation}
a=a_{*}-8\pi^2 T^2 M^3\left(1+\alpha\xi_{2}\right)+\mathcal{O}\left(T^3,\alpha^2\right)\, .
\end{equation}
This relation, the angular velocity, and the extremality condition can be obtained analytically from the results in \cite{Reall:2019sah}, which allows us to get $\xi_{1}$, $\xi_{2}$ and $\xi_{3}$. However, it does not seem possible to obtain the scaling of the time coordinate in \req{newrules} unless one performs explicitly the near-horizon limit of the extremal solutions, which are not yet known.  Therefore, the determination of the parameter $\xi_{4}$ is probably a non-trivial problem. However, as we show below, this parameter ends up playing no role in the near-horizon limit of \req{eq:Twithcorrections2}. 
We also remark that in the relation between $r$ and $\rho$ in \req{newrules}, we defined the coordinate $\rho$ so that $\hat{\Delta}$ takes the right form
\begin{equation}
\hat{\Delta}\to [1+\alpha(\xi_1+\xi_{2}-2\xi_{4})]\epsilon^2\rho(\rho+\rho_{0})\, ,
\end{equation}
when $\epsilon\to 0$.
Finally, we rescale our radial variables,
\begin{equation}\label{Rstilde}
\hat{R}_{s}=[1-\alpha(\xi_{1}+\xi_{2}-2\xi_{4})]\tilde{R}_{s}\,,
\end{equation}
to ensure that the resulting equation has the standard normalization as in \req{Tnext}.

On the other hand, we have to take into account that $\delta \hat{V}_{s}$ is a function of $r$, $\omega$ and $a$, while we consider the mass $M$ to be fixed, 
\begin{equation}
\delta \hat{V}_{s}=\delta \hat{V}_{s}(r, \omega, a)\, .
\end{equation}

By plugging \req{newrules}, \req{newrules2} and \req{Rstilde} into \req{eq:Twithcorrections2} and expanding at first order in $\alpha$ and to quadratic order in $\epsilon$, we obtain 
 \begin{align}\notag 
& \frac{1}{\epsilon^2}\left[\hat\Delta^{-s+1}\frac{d}{dr}\left[\hat\Delta^{s+1}\frac{d\hat{R}_s}{dr}\right] + \left(V_{s}+\alpha \delta \hat{V}_{s}\right)\hat{R}_{s}\right]=\alpha \frac{\tilde{R}_{s}}{\epsilon^2}\bigg[\frac{1}{2} \xi _1 M^2 \left(3 m^2+8 i m s+4 s-4 B_{lm}\right)\\\notag
&+\delta \hat{V}_{s}\bigg]+\alpha\frac{\tilde{R}_{s}}{2\epsilon}\Big[2M\xi _1 \left(2 i m \left(2 \rho +\rho _0\right) s+\om  (m+4 i s)\right)+2\xi_{3} m M ((m-is) (2 \rho+\rho_0)+2 \om)\\\notag
&+(2\rho+\rho_0)\delta \hat{V}_{s}{}^{(1,0,0)}+\frac{\om}{M^2}\delta \hat{V}_{s}{}^{(0,1,0)}\Big]\\\notag
&+\bar{\Delta}^{1-s}\frac{d}{d\rho}\left[\bar{\Delta}^{s+1}\frac{d\tilde{R}_{s}}{d\rho}\right]+\left[\om^2+\frac{\rho_{0}^2m}{4}\left(m-2is\right)+(2\rho+\rho_0)(m- is)\om\right.\\\notag
&\left.+\bar{\Delta}\left(s+\frac{7m^2}{4}-B_{lm}\right)+\alpha\xi _1 \left(\frac{1}{8} m^2 \left(4 \rho ^2+4 \rho _0 \rho -\rho _0^2\right)+\frac{1}{2} \left(2 \rho +\rho _0\right) \om  (m+3 i s)+\frac{\om^2}{2}\right)\right.\\\notag
&\left.+\frac{\alpha}{4} \xi_{3} m \left(m \left(14 \rho^2+14 \rho \rho_0+3 \rho_0^2\right)+8 \om (2 \rho+\rho_0)-2 i \rho_0^2 s\right)\right.\\\notag
&\left.-\frac{1}{2} \xi_{2} \om  \left(2 \om +\left(2 \rho +\rho _0\right) (m-i s)\right)
+\frac{1}{8} \bigg(\frac{\om^2 \delta \hat{V}_s{}^{(0,2,0)}}{M^4}+\frac{2 \left(2 \rho +\rho _0\right) \om \delta \hat{V}_s{}^{(1,1,0)}}{M^2}\right.\\
&\left.+\left(2 \rho +\rho _0\right){}^2 \delta \hat{V}_s{}^{(2,0,0)}-\frac{\rho_0^2}{M} \delta\hat{V}_{s}^{(0,0,1)}\Bigg)\right]\tilde{R}_{s}\, ,
\end{align}
where all the instances of $\delta \hat{V}_{s}$ and its derivatives are evaluated at $(r,\omega, a)=(M,m \Omega_{\rm H}, M)$. Observe, as we mentioned before, that this expression does not depend on the undetermined parameter $\xi_{4}$. 
Now, in order for the near-horizon limit to be well-defined, the $\epsilon^{-2}$ and $\epsilon^{-1}$ terms must vanish. On the other hand, the constant term is the near-horizon limit of the corrected Teukolsky equation, and it must equal the Teukolsky equation of the near-horizon geometry \req{Tnext}.  These conditions allow us to read off all the derivatives of the potential that appear in the expression above. Explicitly, we find\footnote{We recall that our definition of $\delta B_{lm}$ contains the higher-derivative coupling $\alpha$ in it, so the ratio $\delta B_{lm}/\alpha$ is a number independent of $\alpha$, like the coefficients $\xi_i$. }
\begin{equation}\label{deltaVconstr}
\begin{aligned}
\delta \hat{V}_s{}\left(M,m\Omega_{\rm H}, M\right)=&\frac{1}{2} \xi _1 M^2 \left(4 B_{lm}-3 m^2-8 i m s-4 s\right)\,,\\
\delta \hat{V}_s{}^{(1,0,0)}\left(M,m\Omega_{\rm H}, M\right)=&-2 m M \left(\xi _3 (m-i s)+2 i \xi _1 s\right)\,,\\
\delta \hat{V}_s{}^{(0,1,0)}\left(M,m\Omega_{\rm H}, M\right)=&-2 M^3 \left(2 \xi _3 m+\xi _1 (m+4 i s)\right)\,,\\
\delta \hat{V}_s{}^{(2,0,0)}\left(M,m\Omega_{\rm H}, M\right)=&-m^2\left(\xi _1+7 \xi _3\right)-2 \delta B_{lm}/\alpha\,,\\
\delta \hat{V}_s{}^{(1,1,0)}\left(M,m\Omega_{\rm H}, M\right)=&-2 M^2 \left(4 \xi _3 m+\xi _1 (m+3 i s)-\xi_{2} (m-i s)\right)\,,\\
\delta \hat{V}_s{}^{(0,2,0)}\left(M,m\Omega_{\rm H}, M\right)=&-4 M^4 \left(\xi_1-2 \xi_{2}\right)\,,\\
\delta \hat{V}_s{}^{(0,0,1)}\left(M,m\Omega_{\rm H}, M\right)=&-M \left(2 \delta B_{lm}/\alpha+2 \xi _1 m^2+\xi _3 m (m+4 i s)\right)\, .
\end{aligned}
\end{equation}
Using the results of \cite{Reall:2019sah} we find the following values of the $\xi_{1,2,3}$ coefficients for the even parity theories
\begin{align}
\xi_{1}^{(\rm ev)}&= -\frac{1}{7}\,, &\quad \xi_{2}^{(\rm ev)}&= \frac{19}{7}\, ,& \quad \xi_{3}^{(\rm ev)}&= -\frac{19}{28}\, , \\
\xi_{1}^{(\rm 1)}&= \frac{152}{5}\,,&\quad \xi_{2}^{(\rm 1)}&= \frac{3}{4} (1216+315 \pi )\, ,&\quad \xi_{3}^{(\rm 1)}&= \frac{304}{5}\, ,\\
\xi_{1}^{(\rm 2)}&= \frac{148}{5}\,,&\quad \xi_{2}^{(\rm 2)}&= \frac{21}{4} (176+45 \pi )\, ,& \quad \xi_{3}^{(\rm 2)}&= \frac{296}{5}\, ,
\end{align}
while for the odd-parity theories all these coefficients vanish.

 Thus, the relations \req{deltaVconstr} determine the behavior of the potential $\delta \hat{V}_{s}$ [defined by \req{eq:Twithcorrections2}] in the regime of small $(r-M)$, $\omega-m\Omega_{\rm H}$ and $a-M$. By extension, they also fix the form of $\delta V_{s}$ [entering in the equation \req{eq:Twithcorrections}] after taking into account  \req{dVtransf}.
%%%%%%%%%%%%%%%%%%%%%%%%%%%%%%%%%%%%
%%%%%%%%%%%%%%%%%%%%%%%%%%%%%%%%%%%%
\section{Conclusion} \label{sec:conclusion}
%%%%%%%%%%%%%%%%%%%%%%%%%%%%%%%%%%%%
%%%%%%%%%%%%%%%%%%%%%%%%%%%%%%%%%%%%
	
In this work, we have taken the first steps towards analyzing the Teukolsky equation for highly spinning black holes, \textit{i.e.}, near or at extremality, in an extension of Einstein gravity with higher-derivative corrections.  Following the perspective of EFT, we considered an effective action with up to eight-derivative terms, including both parity-preserving and parity-violating terms. Our progress stems from placing our attention to  the near-horizon (near-)extremal geometry of the black holes, which can be found analytically along with many of its properties. 

We then have studied gravitational perturbations by finding modified Teukolsky equations following the approach of \cite{Cano:2023tmv}. These equations become decoupled once one expresses all the perturbed variables in terms of the Teukolsky variables $\delta\Psi_{0,4}$, which we achieved by reconstructing the metric perturbation as a function of $\delta\Psi_{0,4}$. The resulting equations are automatically separable into radial and angular components, and the main result is a modification of the angular separation constants, $\delta B_{lm}$, introduced by the deformation of the angular equation. The final result for $\delta B_{lm}$, given in \eqref{deltaBf}, is an integration of spin-weighted spheroidal harmonics multiplied by a theory-dependent function that we have found explicitly for all theories. The expressions are lengthy and we provide them in an ancillary file. 

A few remarks are now in order. First, we have checked that the corrections to the angular separation constants are, as expected, independent of all gauge conditions. In particular they are independent of the gauge freedom in the background metric \eqref{eq:gaugeh} and of the Starobinsky-Teukolsky constants \req{STidentities}, which represent a gauge choice in the form of the metric perturbation.  Second, unlike the case of GR, the results depend on the polarization of the perturbation, which is characterized in terms of two parameters that we denote by $q_{s}$ --- they are defined in \req{conjugaterelation}. These parameters are fixed by demanding a consistency condition between the different Teukolsky radial equations --- summarized by \req{equalB} --- which always leads to two solutions for the polarization $q_{s}^{\pm}$, naturally corresponding to the two degrees of freedom of the graviton. Each of these families of modes receives in general different corrections to the angular separation constants, denoted by $\delta B_{lm}^{\pm}$. 
We note that all of this is completely analogous to the correction to the quasinormal mode frequencies studied in \cite{Cano:2023tmv,Cano:2023jbk} for non-extremal black holes. 
In particular, the fact that the corrections are different for each polarization is equivalent to the breaking of isospectrality \cite{Li:2023ulk}. 

We have employed numerical and analytical techniques to obtain the corrections to the angular separation constants for both polarizations for various values of $l$ and $m$. For the even-parity theories, the values of $\delta B_{lm}^{+}$ and $\delta B_{lm}^{-}$ are in general independent, while for the odd-parity theories one always finds $\delta B_{lm}^{-}=-\delta B_{lm}^{+}$, so the magnitude $|\delta B_{lm}^{\pm}|$ is polarization-independent. The modes with $m=0$ can be obtained analytically and we show a few of them in  Table~\ref{table:axisymmetrictable}, while $m\neq 0$ modes can only be accessed numerically --- see Table~\ref{table:numericaltable} and Figure~\ref{fig:evenplots}. 
Finally, in the eikonal limit, analytical results can be obtained by taking advantage of certain gauge freedom to drastically simplify the integrand of \req{deltaBf}. The full expressions of $\delta B_{lm}$, provided in Appendix~\ref{app:eikonal}, show that the relative corrections grow as $\delta B_{lm}/B_{lm}\sim l^{0}$ for the cubic theories and as $\delta B_{lm}/B_{lm}\sim l^{2}$ for the quartic ones, when $l\rightarrow \infty$. Therefore, the EFT breaks down in the eikonal limit for the quartic theories --- it will happen when $l\sim |\alpha_{\rm q}|^{-1/2}$, where $\alpha_{\rm q}$ is one of the couplings defined in \req{alphaq} --- but surprisingly, it does not break down in the case of the cubic theories, since the relative corrections are bounded.

We then discussed some implications of the corrected radial Teukolsky equations at extremality \req{eq:RadialCorrected} and near-extremality \req{Tnext}. 
We have connected our results with those of \cite{Horowitz:2023xyl}, that found that static and axisymmetric perturbations of the near-horizon geometry give rise to curvature singularities in the presence of higher-derivative corrections. This reference focused on the even-parity cubic and quartic theories and made use of metric perturbations to reach their conclusions.  We then reexamined their results in the light of the corrected Teukolsky equation, which on the other hand allows us to consider non-static and non-axisymmetric perturbations as well.  
As we explained in Section~\ref{sec:divergence}, our results precisely match those of \cite{Horowitz:2023xyl} for the even-parity theories. In particular, we find that the Teukolsky variable $\Psi_0$ behaves as $\Psi_{0}\sim \rho^{\gamma}$, with a theory-dependent exponent $\gamma$ that agrees with those given in \cite{Horowitz:2023xyl}. For the $l=2$ mode, one finds $\gamma=\mathcal{O}(\alpha)$ and divergences occur for the signs of the higher-derivative coupling constants such that $\gamma<0$. 
However, \cite{Horowitz:2023xyl} did not take into account parity-breaking corrections, as it was stated there that those do not affect the result. Here, we have found that parity-breaking corrections are indeed relevant, as they also give rise to a correction to the exponent $\gamma$. In particular, for the odd-parity quartic theory we found the result \req{gammaquartic3}, which always gives rise to a divergence of $\Psi_0$ for one of the polarization modes.  For the odd-parity cubic theory we found $\gamma=0$ instead, but this seems to be an accident since $\gamma$ is non-zero for other values of $l$ different from 2. 
We remark that the study of perturbations of parity-breaking theories is particularly intricate as the correction to $\gamma$ comes from the effect of coupling modes of odd and even parity together. If one of the modes is (inconsistently) truncated, one would get the wrong result that the corrections vanish.

Our final remarks are on the connection between the near-horizon near-extremal limit of the full radial Teukolsky equation and the Teukolsky equation obtained directly from the near-horizon geometry. By using appropriate changes of variable, the corrections to the full Teukolsky equation can always be written in the form \req{eq:Twithcorrections2} which depends on a correction to the potential $\delta \hat{V}_s$. 
Determining $\delta \hat{V}_s$ would allow one to compute the QNM frequencies of the corrected near-extremal black holes, but the problem is that this potential is unknown for black holes close to extremality. In fact, $\delta \hat{V}_s$ [or more precisely, its non-hatted version \req{dVtransf}] has so far only been found for non-extremal black holes as a series expansion in the angular momentum \cite{Cano:2023jbk} which breaks down at extremality. 
Our results here allow us to make substantial progress in the determination of $\delta \hat{V}_s$ by establishing precise constraints on the form of $\delta \hat{V}_s$ in the near-horizon region of near-extremal black holes --- see Eq.~\req{deltaVconstr}. 
Our hope would be to combine these constraints together with input on behavior of $\delta \hat{V}_s$ in the asymptotic region, which would allow us to fix the full potential unambiguously. 
This would allow one to estimate the corrections to the QNM frequencies of the ``zero damping modes'' \cite{Yang:2013uba} --- those that approach $\omega\sim m \Omega_{\rm H}$ for extremal Kerr.   We expect to report on this elsewhere.

\section*{Acknowledgements}

The work of PAC received the support of a fellowship from “la Caixa” Foundation (ID 100010434) with code LCF/BQ/PI23/11970032. MD is supported in part by the Odysseus grant (G0F9516N Odysseus) as well as from the Postdoctoral Fellows of the Research Foundation - Flanders grant (1235324N).

\appendix
\section{Universal Teukolsky equations}\label{app:Teukolsky}
Here we show the form of the univeral Teukolsky equations \cite{Cano:2023tmv} that we use to obtain \req{eq:modTeukPsi}. These equations are best written in the Geroch-Held-Penrose (GHP) form of the Newman-Penrose formalism, so let us briefly introduce the basic quantities. 

Given a NP frame with four null vectors $\{l_{\mu},n_{\nu},m_{\mu},\bar{m}_{\mu}\}$ satisfying
\begin{equation}
g_{\mu\nu}=-2l_{(\mu}n_{\nu)}+2m_{(\mu}\bar{m}_{\nu)}\, ,
\end{equation}
 the different components of the spin connection are denoted by
\begin{equation}\label{eqn:spincoefficients}
	\begin{aligned}
\kappa &= -m^{\mu}l^{\nu}\nabla_{\nu}l_{\mu} \, , \qquad 
\sigma = -m^{\mu}m^{\nu}\nabla_{\nu}l_{\mu} \, , \qquad 
 \sigma'  =n^{\mu}\bar{m}^{\nu}\nabla_{\nu}\bar{m}_{\mu} \, , \qquad
\kappa' = n^{\mu}n^{\nu}\nabla_{\nu}\bar{m}_{\mu}  \, , \\ \\ 
\rho	 &= -m^{\mu}\bar{m}^{\nu}\nabla_{\nu}l_{\mu} \, ,  \qquad
\rho' = n^{\mu}m^{\nu}\nabla_{\nu}\bar{m}_{\mu} \, ,  \qquad
\tau	 = -m^{\mu}n^{\nu}\nabla_{\nu}l_{\mu} \, , \qquad
\tau' = n^{\mu}l^{\nu}\nabla_{\nu}\bar{m}_{\mu} \, ,  \\ \\
\epsilon &= -\frac{1}{2}(n^{\mu}l^{\nu}\nabla_{\nu}l_{\mu} + m^{\mu}l^{\nu}\nabla_{\nu}\bar{m}_{\mu}  ) \, , \qquad 
\epsilon' = \frac{1}{2}(n^{\mu}n^{\nu}\nabla_{\nu}l_{\mu} + m^{\mu}n^{\nu}\nabla_{\nu}\bar{m}_{\mu}) \, , \\ \\
\beta' &= \frac{1}{2}(n^{\mu}\bar{m}^{\nu}\nabla_{\nu}l_{\mu}+ m^{\mu}\bar{m}^{\nu}\nabla_{\nu}\bar{m}_{\mu}) \, ,\qquad  \beta = -\frac{1}{2}(n^{\mu}m^{\nu}\nabla_{\nu}l_{\mu} + m^{\mu}m^{\nu}\nabla_{\nu}\bar{m}_{\mu}) \, . 
	\end{aligned}
\end{equation}
We also have the components of the Weyl tensor, denoted by
\begin{equation}\label{eqn:appintro:curv2} 
	\begin{aligned}
\Psi_0 &= C_{\alpha \beta \mu \nu} l^{\alpha}m^{\beta}l^{\mu}m^{\nu} \, , \qquad \Psi_1 = C_{\alpha \beta \mu \nu} l^{\alpha}n^{\beta}l^{\mu}m^{\nu} \, ,  \qquad
\Psi_2 = \Psi_2' = C_{\alpha \beta \mu \nu} l^{\alpha}m^{\beta} \bar m^{\mu}n^{\nu}\, , \\ \\ \qquad \Psi_3 &= \Psi_1' = C_{\alpha \beta \mu \nu} l^{\alpha}n^{\beta}\bar m^{\mu} n^{\nu} \, ,  \qquad \Psi_4 = \Psi_0' = C_{\alpha \beta \mu \nu} n^{\alpha}\bar{m}^{\beta}n^{\mu}\bar{m}^{\nu} \, .
	\end{aligned}
\end{equation}
Observe that the prime $'$ denotes the simultaneous exchange 
\begin{equation}\label{primedef}
':\quad l_{\mu}\leftrightarrow n_{\mu}\, , \quad m_{\mu}\leftrightarrow \bar{m}_{\mu}\, .
\end{equation}

For compactness, we will introduce the GHP derivatives, which are covariant under local rescalings of the form
	\begin{equation}\label{eqn:intro:typeIII}
	l^\mu \mapsto e^{\lambda}l^\mu,\qquad n^\mu \mapsto  e^{-\lambda} n^\mu,\qquad  m^\mu \mapsto e^{i \theta} m^\mu\qquad \bar m^\mu \mapsto e^{-i \theta} \bar m^\mu \, .
\end{equation}
where $\lambda,\theta \in \mathbb R$. We say that an object $X$ transforms with weight $w_{\rm GHP}(X) = \{p,q\}$ if it transforms under the change of frame \eqref{eqn:intro:typeIII} as
\begin{equation}
 X \mapsto e^{\frac{p}{2}\left(\lambda+i \theta\right)+\frac{q}{2}\left(\lambda-i \theta\right)}X \, .
\end{equation}
The weights of the frame fields are
	\begin{equation}\label{eqn:intro:typeIIIweights}
	w_{\rm GHP}(l^\mu)=  \{1,1\} \, ,\quad w_{\rm GHP}(n^\mu)=  \{-1,-1\} \, ,\quad  w_{\rm GHP}(m^\mu)=  \{1,-1\} \, ,\quad  w_{\rm GHP}(\bar{m}^\mu)=  \{-1,1\} \, .
\end{equation}
while the spin coefficients with definite weight are
\begin{equation}
		w_{\rm GHP}(\kappa)=  \{3,1\} \, ,\quad w_{\rm GHP}(\sigma)=  \{3,-1\} \, ,\quad  w_{\rm GHP}(\rho)=  \{1,1\} \, ,\quad  w_{\rm GHP}(\tau)=  \{1,-1\} \, ,
\end{equation}
On the other hand, $\epsilon$, $\epsilon'$, $\beta$, $\beta'$ are used to construct the GHP derivatives of definite weight, which take the form
\begin{equation}\label{eqn:GHPderivatives}
	\begin{aligned}
		\th   = (l^\alpha \nabla_\alpha - p \epsilon - q \epsilon^*),  \quad
		\th'  = (n^\alpha \nabla_\alpha + p  \epsilon' + q \epsilon'{}^*), \\ \\
		\edth  = (m^\alpha \nabla_\alpha - p \beta + q \beta'{}^*), \quad
		\edth' = (\bar{m}^\alpha \nabla_\alpha + p \beta' - q \beta^*) \, ,
	\end{aligned}
\end{equation} 
when acting on an object of weight $\{p,q\}$.

Now we are ready to state the form of the universal Teukolsky equation. For a theory satisfying the equations of motion
\begin{equation}
G_{\mu\nu}=T_{\mu\nu}\, ,
\end{equation}
the universal Teukolsky equation for $\Psi_0$ reads
\begin{equation}\label{eqn:universalteukolsky0}
\cO^{(0)}_{+2}\left(\Psi_0\right) +\cO^{(i)}_{+2}\left(\Psi_1\right)+ \cO^{(ii)}_{+2}\left(\Psi_0\right)   = \cT^{(0)}_{+2} +\cT^{(i)}_{+2} +\cT^{(ii)}_{+2} \, ,
\end{equation}
with\footnote{The operator $\cO^{(i)}_{+2}$ should not be confused with the operator $\mathcal{O}_{s}^{(1)}$ appearing in equation \req{eq:modTeukPsi}, which is the linearized version of \req{eqn:universalteukolsky0}. }
\begin{subequations}
	\bea\label{TeukOpDef}
	\cO^{(0)}_{+2} &=& 2 \left[(\th - 4 \rho - \rho^*)(\th'-\rho') - (\edth-4\tau-\tau'{}^*)(\edth'-\tau') -3\Psi_2\right] \, , \\
	\cO^{(i)}_{+2} &=& 4 \left[2\kappa\left(\th'-\rho'^*\right)-2\sigma\left(\edth'-\tau^*\right)+2\left(\th'\kappa\right)-2\left(\edth'\sigma\right)+5\Psi_1 \right] \, , \\
	\cO^{(ii)}_{+2} &=& 6 \left[\kappa \kappa'-\sigma \sigma'\right] \, , \\  \nn
	\cT^{(0)}_{+2}  &=& (\edth -\tau'{}^*-4\tau)[(\th-2\rho^*)T_{lm}-(\edth -\tau'{}^*)T_{ll}] \nn &+& (\th-4\rho-\rho^*)[(\eth - 2\tau'{}^*)T_{lm}-(\th-\rho^*)T_{mm}] \, , \\
	\cT^{(i)}_{+2}  &=& \frac{1}{2}\left[\sigma \th -\kappa \edth \right]T - \left[3 \sigma \left(\th' -\rho'{}^*\right) -\sigma'{}^*\left(\th -4\rho-\rho^*\right)- \th\left(\sigma'{}^*\right)\right]T_{ll} \nn 
	&-& 2\left[\sigma \left(\edth -\tau-\tau'{}^*\right) +\edth\left(\sigma\right) \right] T_{l\bar{m}}+\left[3\sigma \left(\edth' -2\tau{}^*\right) +3\kappa\left(\th' -2\rho'{}^*\right)\right] T_{lm} \nn 
	&-& \left[3\kappa \left(\edth'-\tau^*\right) - \kappa^* \left(\edth-4\tau-\tau'{}^*\right) - \edth\left(\kappa^*\right) \right] T_{mm} \nn 
	&+& \left[\kappa \edth + \sigma \left(\th-2\rho-2\rho^*\right) +2 \th\left(\sigma\right)-\Psi_0 \right] \left(T_{ln}+T_{m\bar{m}}\right) \nn 
	&-& 2\left[\kappa \left(\th -\rho-\rho^*\right) + \th\left(\kappa\right)\right] T_{nm} \, , \\
	\cT^{(ii)}_{+2}  &=& 3\left[\kappa \kappa'{}^*T_{ll}+\sigma \sigma{}^* T_{mm}\right] \, , 
	\eea
\end{subequations}
where $T_{ll}=T_{\mu\nu}l^{\mu}l^{\nu}$, $T_{ln}=T_{\mu\nu}l^{\mu}n^{\nu}$ and so on, and $T=T_{\mu\nu}g^{\mu\nu}$. 
On Ricci flat, Petrov-D background, this reduces to the usual Teukolsky equation for the linear fluctuations of $\Psi_0$,
\begin{equation}
\cO^{(0)}_{+2}\left(\Psi_0\right)=0\, .
\end{equation}
The equation for $\Psi_{4}$ is obtained by applying the prime conjugation \req{primedef} to \req{eqn:universalteukolsky0}, while those for $\Psi_{0}^{*}$ and $\Psi_{4}^{*}$ are obtained from the NP conjugation of the $\Psi_0$ and $\Psi_4$ equations.

\section{Eikonal limit}\label{app:eikonal}
Here we show the explicit expressions for the corrections to the angular separation constants $\delta B_{lm}$ in the eikonal limit as a result of the integration of \req{magicformula}. Let us write
\begin{equation}
\delta B_{lm}=\alpha_{\rm ev} \delta B_{lm}^{(\rm ev)}+\alpha_{\rm odd} \delta B_{lm}^{(\rm odd)}+\alpha_{\rm 1} \delta B_{lm}^{(\rm 1)}+\alpha_{\rm 2} \delta B_{lm}^{(\rm 2)}+\alpha_{\rm 3} \delta B_{lm}^{(\rm 3)}\, ,
\end{equation}
so that each coefficient $\delta B_{lm}^{(\rm q)}$ corresponds to each correction. We find
\begin{align}\notag
\delta B_{lm}^{(\rm ev)}=&B_{lm}\Bigg[\frac{\left(15 x_0^8-276 x_0^6-566 x_0^4+1036 x_0^2+31\right) \Pi \left(-x_0^2,-k\right)}{16
   \left(x_0^2+1\right){}^2 \left(x_0^4-x_0^2+4\right) K(-k)}\\\notag
   &+\frac{\left(x_0^2+3\right) \left(139 x_0^8+324
   x_0^6-3494 x_0^4+8980 x_0^2-3725\right) E(-k)}{112 \left(x_0^2+1\right){}^4
   \left(x_0^4-x_0^2+4\right) K(-k)}\\\label{Beveik}
   &+\frac{59 x_0^8+632 x_0^6+2934 x_0^4-9936 x_0^2+4087}{56
   \left(x_0^2+1\right){}^3 \left(x_0^4-x_0^2+4\right)}\Bigg]+i \delta B_{lm}^{(\rm odd)}\, ,\\\notag
\delta B_{lm}^{(\rm odd)}=&\frac{i\, B_{lm}}{q_{2} \left(x_0^4-x_0^2+4\right)}\Bigg[\frac{45  \Pi
   \left(-x_0^2,-k\right)}{64 \left(x_0^2+1\right){}^4  K(-k)} \big(15 x_0^{12}-318 x_0^{10}-1959 x_0^8+17404 x_0^6\\\notag
   &-50799 x_0^4+55074 x_0^2-18457\big)+\frac{3 
   \left(x_0^2+3\right)E(-k)}{448 \left(x_0^2+1\right){}^6
   K(-k)} \left(1541 x_0^{12}+7790 x_0^{10}\dvv -363909 x_0^8-1544956 x_0^6+5509931 x_0^4-4704978
   x_0^2+1193205\right)\\\notag
   & -\frac{3 }{224  \left(x_0^2+1\right){}^5} \left(779 x_0^{12}-5530 x_0^{10}+92409 x_0^8-2465212 x_0^6+5534285 x_0^4\dvvtag-3771402
   x_0^2+713295\right)\Bigg]\, ,\label{Boddeik}
\end{align}
for the cubic theories, while for the quartic theories we have
\begin{align}\notag
\delta B_{lm}^{(\rm 1)}=&(B_{lm})^2\Bigg[\frac{3 \left(73 x_0^4+174 x_0^2+165\right)
   \left(x_0^2+1\right)}{\left(x_0^4-x_0^2+4\right){}^2}-\frac{9 \left(63 x_0^4+86 x_0^2+103\right) \left(x_0^2+1\right){}^2 \Pi \left(-x_0^2,-k\right)}{2
   \left(x_0^4-x_0^2+4\right){}^2 K(-k)}\\ \label{B1eik}
   &-\frac{3 \left(x_0^2+3\right) \left(103 x_0^4+174 x_0^2+135\right)
   E(-k)}{2 \left(x_0^4-x_0^2+4\right){}^2 K(-k)}\Bigg]+i \delta B_{lm}^{(3)}\, ,\\\label{B2eik}
   \delta B_{lm}^{(\rm 2)}=&\delta B_{lm}^{(\rm 1)}-2i \delta B_{lm}^{(3)}\, ,\\\notag
   \delta B_{lm}^{(3)}=&\frac{i\, (B_{lm})^2}{640 q_{2} \left(x_0^2+1\right){}^4 \left(x_0^4-x_0^2+4\right){}^2}\Bigg[\frac{45\Pi \left(-x_0^2,-k\right)}{2 K(-k)} \left(427 x_0^{16}-1224 x_0^{14}-20844 x_0^{12}\dvv
   +34632 x_0^{10}+87970 x_0^8-631288 x_0^6+1066836
   x_0^4-494344 x_0^2+67147\right) \dv
   +\frac{3 \left(x_0^2+3\right)E(-k)}{2 K(-k)}
   \left(7359 x_0^{12}+52706 x_0^{10}-268375 x_0^8-104516 x_0^6+1708513 x_0^4\dvv
   -787902 x_0^2-136809\right)
   -3 \left(x_0^2-3\right) \left(3441 x_0^{12}+23198 x_0^{10}-34937 x_0^8-133692
   x_0^6\dvvtag-1220689 x_0^4+950206 x_0^2-58503\right)\Bigg]\, .\label{B3eik}
\end{align}
Here $K(-k)$ is the elliptic integral of the first kind \req{ellipk}, while $E$ and $\Pi$ are the elliptic integrals of the second and third kind, respectively, 

 \begin{align}\
 E(-k)&=\int_{0}^{\pi/2} d\theta \sqrt{1+k \sin^2(\theta)}\, ,\\
 \Pi(-x_0^2,-k)&=\int_{0}^{\pi/2} d\theta  \left[1+x_0^2 \sin^2(\theta)\right]^{-1}\left[1+k \sin^2(\theta)\right]^{-1/2}\, ,
 \end{align}
 and we recall that $k$ is given by \req{kdef}.
		
	%\renewcommand{\leftmark}{\MakeUppercase{Bibliography}}
	%\phantomsection
	\bibliographystyle{JHEP}
	\bibliography{Gravities.bib}
	%\label{biblio}
	
\end{document}